\documentclass[%
 reprint,
 amsmath,amssymb,
 aps,
prb,
]{revtex4-2}

\usepackage{graphicx}
\usepackage{dcolumn}
\usepackage{bm}

\begin{document}

\preprint{APS/123-QED}


\title{Learning thermodynamics and topological order of the 2D XY model with generative real-valued restricted Boltzmann machines}

\author{Kai Zhang}
 \email{kzhang@uttyler.edu}
\affiliation{%
Department of Chemistry and Biochemistry,
The University of Texas at Tyler, Tyler, TX 75799, USA
}%
\date{\today}

\begin{abstract}
Detecting the topological Kosterlitz–Thouless (KT) transition in the prototypical 2D XY model using unsupervised machine learning methods has long been a challenging problem due to the lack of suitable order parameters. To address this issue, we begin with a conventional real-valued RBM (RBM-xy), which uses exponential conditional probabilities to generate visible units. We then develop a novel real-valued RBM (RBM-CosSin) featuring nonlinear $\cos$/$\sin$ activation, whose visible units follow the von Mises distribution. Our findings reveal that RBM-CosSin effectively learns the underlying Boltzmann distribution of 2D XY systems and generate authentic XY configurations that accurately capture both thermodynamics and topological order (vortex). Furthermore, we demonstrate that it is possible to extract phase transition information, including the KT transition, from the weight matrices without relying on prior physics knowledge. 
\end{abstract}

\maketitle


\section{Introduction}
Building on a decade of success in many areas, machine learning (ML) is emerging as a powerful tool for studying phase transitions in condensed matter physics~\cite{carleo2019,carrasquilla2020}. In the presence of symmetry breaking, it is quite compelling that ML algorithms can extract useful information about order parameters and transition temperatures from raw state configurations without physics knowledge~\cite{carrasquilla2017,wetzel2017}.  As the simplest system with symmetry-broken phases, the Ising model has discrete configuration space that can be effectively learned by various machine learning models~\cite{torlai2016,tanaka2017,funai2020,DAngelo2020}.
Even for the three-dimensional (3D) XY model with continuous symmetry, extraction of order parameters with variational autoencoders (VAEs) and clustering of different phases with t-distributed stochastic neighbor embedding (TSNE) is still possible~\cite{wetzel2017}. In all these instances, the low-temperature ordered phases exhibit true long-range order, which gives rise to specific order parameters. 
Beyond this paradigm lies the Kosterlitz–Thouless (KT) transition~\cite{kosterlitz1973}. 

Unlike the ferromagnetic-paramagnetic transition in the Ising and 3D XY models, the KT transition is topological in nature and lacks a clearly identifiable order parameter. 
Accurately capturing topological order is the prerequisite  for the quantitative analysis of systems undergoing KT transitions.
In the prototypical two-dimensional (2D) XY model, the KT transition is driven by the unbinding of vortex-antivortex pairs~\cite{kosterlitz1973}. Supervised learning, in which the phases of configurations are labeled, can perform phase classification task if topological defects (vortex) are provided by human researcher or learned from raw spin states using customized neural networks, such as convolutional neural networks and graph-convolution networks~\cite{beach2018,zhang2019,song2022}. The difficulty in discriminating 2D XY phases through supervised learning raises concerns about the  challenge in unsupervised learning of KT transition from unlabeled data~\cite{beach2018}. 

Early efforts of unsupervised learning of the 2D XY model involve linear or kernel principal component analysis (PCA) of the $x,y$-components XY spins~\cite{hu2017,wang2017,wang2018}. Although clustering of data points of high and low temperature ``phases'' on the low dimensional principal subspace is observed, it is not related to the KT transition. In fact, PCA cannot even recognize the existence of vortices and this apparent phase discrimination corresponds to a crossover temperature above the KT transition. Facing this failure, it was believed that an automated ML algorithm cannot be expected to learn about unbinding of vortex-antivortex pairs from raw spin configurations~\cite{hu2017}. Meanwhile, various versions of VAEs are employed to extract latent variables that represent magnetization and susceptibility. However, these variables are misinterpreted as predictors of the KT transition, when they actually point to the higher crossover temperature~\cite{cristoforetti2017,naravane2023}.
   
Recently, advanced unsupervised learning methods based on diffusion maps and clustering algorithms start to make quantitative predictions about the KT transition by analyzing the average cluster distance and within-cluster dispersion~\cite{rodriguez2019}.
A similar study suggests using the maximum of the Calinski-Harabaz (ch) index as an indicator of the KT transition~\cite{wang2021}. Aside from the conceptual complexities, these unsupervised learning methods do not learn the underlying Boltzmann distribution of XY states, making them incapable of data generation. So far, no generative models have demonstrated the ability to satisfactorily learn the thermodynamic properties (such as energy and heat capacity) and vortex behavior of 2D XY configurations~\cite{cristoforetti2017}.

To tackle these challenges, we resort to  classical restricted Boltzmann machines (RBMs), which are two-layer energy-based models~\cite{leRoux2008, fischer2012, hinton2012}. The standard binary-unit RBMs have effectively addressed the Ising phase transition~\cite{torlai2016,morningstar2018,iso2018,cossu2019,DAngelo2020,funai2020}.
Given that XY spins are continuous, a natural extension is to use RBMs with real-valued visible units (hidden units can still be binary). Previously, approximately continuous RBMs or unrestricted Boltzmann machines have been utilized in studies of XY spin glasses with multi-valued units~\cite{lin1995}, as well as in lattice protein models using categorical variables~\cite{tubiana2019}.
The most commonly used real-valued RBMs have Gaussian units~\cite{fischer2012}, which are unsuitable for bounded XY spin values. 

In this article, we propose two types of real-valued RBMs (RBM-xy and RBM-CosSin) that learn from equilibrium XY configurations obtained by Monte Carlo simulations and investigate their abilities to generate new Boltzmann-distributed configurations. We find that both RBMs can genuinely capture thermodynamic properties and topological order of the 2D XY model, with RBM-CosSin demonstrating superior performance. By analyzing the weight matrices of RBM-CosSin, we show that important information about the KT transition and the crossover transition (at a higher temperature) can be extracted from ML parameters alone. Below, we first introduce our XY datasets and RBMs in Section~\ref{sec:method}. In Section~\ref{sec:result}, we present and discuss our findings on RBM training, the generation of new XY configurations, the capture of thermodynamic properties, the capture of topological order (vortex), and the analysis of the weight matrix. A conclusion follows in Section~\ref{sec:conclusion}.

\section{Models and Methods}
\label{sec:method}

\subsection{2D XY configurations generated by Monte Carlo simulations}
We consider the 2D XY model on square lattices with $N=L\times L$ spins ($L=16,32$) under periodic boundary conditions. The Hamiltonian of the system with ferromagnetic coupling  ($J>0$) between nearest neighbors $\langle i,j \rangle$ is
\begin{align}
H_{XY} = - J \sum_{\langle i,j\rangle} {\bf s}_i \cdot {\bf s}_j = - J \sum_{\langle i,j\rangle} \cos (\varphi_i - \varphi_j),
\end{align}
where ${\bf s}_i = (x_i, y_i)=(\cos \varphi_i, \sin \varphi_i)$ is the vector spin  variable and $\varphi_i\in [0,2\pi]$ is the angle of ${\bf s}_i$ at each site $i=1,2,\dots,N$. We generate $20000$ equilibrium configurations at each of the sixteen temperatures in the range of $T=0.6$-$1.7$ (in units of $J/k_B$ with $k_B$ being the Boltzmann constant) using standard Monte Carlo simulations. The internal energy $\langle E\rangle$, heat capacity $C_V = (\langle E^2\rangle - \langle E \rangle^2)/(k_B T^2)$, magnetization, 
\begin{align}
\begin{aligned}
\langle  |{\bf M}| \rangle &= \left\langle  \left|\sum\limits_i   {\bf s}_i \right|\right\rangle \\
&=\left\langle \sqrt{\left(\sum_i \cos \varphi_i\right)^2 + \left(\sum_i \sin \varphi_i \right)^2 }\right\rangle
\end{aligned}
\end{align}
and susceptibility $\chi =  \left(\langle  {\bf M}^2 \rangle - \langle  |{\bf M}| \rangle^2 \right)/(k_B T)$ are measured (in reduced units) and to be compared with ML results later. 

The spontaneous magnetization of the 2D XY model should vanish in the thermodynamic limit, i.e. $\langle  |{\bf M}| \rangle \to 0$ as $N\to \infty$. One can use the mean-squared magnetization 
\begin{align}
\langle  {\bf M}^2 \rangle  =   \left\langle \left(\sum_i \cos \varphi_i\right)^2 + \left(\sum_i \sin \varphi_i \right)^2 \right\rangle
\end{align}
as an approximation to $k_BT \chi$.

Early Monte Carlo simulations of the 2D XY model observed a
system size independent peak of $C_V/N$ at a height of about 1.5 around a crossover temperature $T_p = 1.02(5)$~\cite{tobochnik1979, ota1992} (the recent estimate is $1.043(4)$~\cite{nguyen2021}. This behavior is related to energy cost of vortex formation~\cite{ota1992}, while the true KT transition occurs at lower temperature $T_c = 0.893(1)$~\cite{komura2012}. The divergence of correlation length and susceptibility (assuming zero magnetization) near $T_c$ fitted  to the KT theory can provide a rough estimate to $T_c$~\cite{tobochnik1979,gupta1988}.  The accurate determination of $T_c$ should be obtained with finite-size scaling of helicity modulus or other advanced quantities~\cite{weber1988,olsson1995,komura2012}, which is beyond the scope of the current study.

\subsection{Restricted Boltzmann machines with real-valued visible units (RBM-xy)}
We first consider restricted Boltzmann machines (RBMs) with $n_h$ binary hidden units, $h_i \in \{0,1\}$ ($i=1,2,\cdots, n_h$),  and $n_v$ real-valued visible units, $v_j\in[-1,1]$ ($j=1,2,\cdots,n_v$).  
The state vectors of the hidden layer and the visible layer are denoted as  ${\bf h} = [h_1, h_2, \cdots, h_{n_h}]^T$ and ${\bf v} = [v_1, v_2, \cdots, v_{n_v}]^T$, respectively. Throughout the article we use the convention that a vector ${\bf r}$ is   a column vector and  its row vector counterpart is explicitly expressed as a transpose ${\bf r}^T$.
To study the 2D XY model with $N$ spins, we define the visible layer vector to be the concatenation of all $x,y$-components ${\bf v}^T = [{\bf x}^T; {\bf y}^T ] \equiv [x_1, x_2, \cdots, x_N; y_1, y_2, \cdots, y_N]$ with 
$n_v = 2N$ units. The ideal of using $x,y$-components to represent spin configurations was widely adopted in other ML studies~\cite{hu2017,wang2017}.  For this reason, the RBM in this section is termed as ``RBM-xy''.

The total energy $E_{\boldsymbol{\theta}}({\bf v}, {\bf h})$ of RBM-xy at an overall state $({\bf v}, {\bf h})$ is
\begin{equation}
\begin{aligned}
E_{\boldsymbol{\theta}}({\bf v}, {\bf h}) &= - {\bf b}^T {\bf v} - {\bf c}^T {\bf h} - {\bf h}^T{\bf W} {\bf v} \\
&= - \sum\limits_{j=1}^{n_v} b_j v_j - \sum\limits_{i=1}^{n_h} c_i h_i -\sum\limits_{i=1}^{n_h} \sum\limits_{j=1}^{n_v} W_{ij} h_i v_j
\end{aligned}
\end{equation}
where ${\bf b} = [b_1, b_2, \cdots, b_{n_v}]^T$ is the visible bias, ${\bf c} = [c_1, c_2, \cdots, c_{n_h}]^T$ is the hidden bias, and
\begin{equation}
{\bf W}_{n_h \times n_v} = 
\begin{bmatrix}
-{\bf w}_1^T-\\
-{\bf w}_2^T-\\
\vdots\\
-{\bf w}_{n_h}^T-\\
\end{bmatrix}
=
\begin{bmatrix}
| & | & & |\\
{\bf w}_{:, 1} & {\bf w}_{:, 2} & \cdots & {\bf w}_{:, n_v} \\
| & | & & |
\end{bmatrix}
\end{equation}
 is the interaction weight matrix between visible and hidden units. Under this notation, a row vector ${\bf w}_i^T$ is a filter mapping from spin configurations to a hidden unit $i$ and a column vector ${\bf w}_{:,j}$ is an inverse filter mapping from   hidden states to a visible unit $j$ (a spin).
All parameters are collectively written as ${\boldsymbol\theta} \equiv \{ {\bf W}, {\bf b}, {\bf c} \}$.

Let $\alpha_i({\bf v}) ={\bf w}_i^T {\bf v} + c_i  $ and $\beta_j({\bf h}) = b_j + {\bf h}^T {\bf w}_{:,j} $ be the conjugate mean field on hidden unit $i$ and visible unit $j$, respectively. They are the components of vectors  ${\boldsymbol \alpha}({\bf v}) = [\alpha_1, \alpha_2, \cdots, \alpha_{n_h}]^T = {\bf W} {\bf v} + {\bf c}$ and  ${\boldsymbol \beta}({\bf h}) = [\beta_1, \beta_2, \cdots, \beta_{n_v}]^T = {\bf b} +  {\bf W}^T {\bf h}$. Under these definitions, the total energy can be rewritten as
\begin{equation}
\begin{aligned}
E_{\boldsymbol{\theta}}({\bf v}, {\bf h}) &= - {\bf b}^T {\bf v} -  {\bf h}^T {\boldsymbol \alpha}({\bf v}) \\
&=   - {\bf c}^T {\bf h} - {\boldsymbol \beta}^T({\bf h}) {\bf v}.
\end{aligned}
\end{equation}

The partition function of the RBM is the (intractable unless for small simple systems) summation/integral of the Boltzmann factor $e^{- E_{\boldsymbol \theta}({\bf v}, {\bf h})}$ over all $({\bf v}, {\bf h})$ states
\begin{equation}
\begin{aligned}
Z_{\boldsymbol \theta} &= \int d{\bf v} \sum_{\bf h} e^{-E_{\boldsymbol \theta}({\bf v}, {\bf h})} \\
&=\int d{\bf v} e^{-{\mathcal E}_{\boldsymbol \theta}({\bf v})} = \sum_{\bf h} e^{-{\mathcal F}_{\boldsymbol \theta}({\bf h})}
\end{aligned}
\end{equation}
where $\int d {\bf v} = \int_{-1}^1 dv_1 \int_{-1}^1 dv_2 \cdots \int_{-1}^1 dv_{n_v} $ and $\sum\limits_{\bf h} = \sum\limits_{h_1=0}^1  \sum\limits_{h_2=0}^1 \cdots  \sum\limits_{h_{n_h}=0}^1$. 
The  visible energy ${\mathcal E}_{\boldsymbol \theta}({\bf v})$ (often termed as free energy in machine learning literature) and hidden energy ${\mathcal F}_{\boldsymbol \theta}({\bf h})$ are defined through 
 $e^{- {\mathcal E}_{\boldsymbol \theta}({\bf v}) }  = \sum\limits_{\bf h}   e^{- E_{\boldsymbol \theta}({\bf v}, {\bf h})} $ and  $e^{- {\mathcal F}_{\boldsymbol \theta}({\bf h}) }  = \int d{\bf v}   e^{- E_{\boldsymbol \theta}({\bf v}, {\bf h})} $. In most RBMs studies, only ${\mathcal E}_{\boldsymbol \theta}({\bf v})$ is useful and here it can be shown that
\begin{equation}
\begin{aligned}
 {\mathcal E}_{\boldsymbol \theta}({\bf v}) &= -{\bf b}^T {\bf v} - \sum\limits_{i=1}^{n_h} \ln \left(1+e^{ {\bf w}_i^T {\bf v} + c_i }\right)\\
 & =  -{\bf b}^T {\bf v} - \sum\limits_{i=1}^{n_h} \ln \left(1+e^{ \alpha_i({\bf v}) }\right).
\end{aligned}
\end{equation}

The model distribution $p_{\boldsymbol \theta}({\bf v})$ for a visible state ${\bf v}$ can be introduced by marginalization of the joint probability distribution  $p_{\boldsymbol \theta}({\bf v}, {\bf h}) = e^{- E_{\boldsymbol \theta}({\bf v}, {\bf h})}/Z_{\boldsymbol \theta}$
\begin{equation}
p_{\boldsymbol \theta}({\bf v}) = \sum_{\bf h} p_{\boldsymbol \theta}({\bf v}, {\bf h}) = \frac{1}{Z_{\boldsymbol \theta}} e^{- {\mathcal E}_{\boldsymbol \theta}({\bf v})}.
\end{equation}
 The training of RBMs involves maximizing this likelihood function $p_{\boldsymbol \theta}({\bf v})$ (or its logarithmic) with respect to parameters ${\boldsymbol \theta}$ given data points ${\bf v}  \in {\mathcal D}$ drawn independently from the identical data distribution $p_{\mathcal D}({\bf v})$. Although $p_{\boldsymbol \theta}({\bf v})$ is not accessible due to the intractable $Z_{\boldsymbol \theta}$, the resulting terms in gradient decent algorithms can be sampled as will be explained next.

 The loss function of RBM learning is the negative log likelihood under maximum likelihood estimation
 \begin{equation}
{\mathcal L}({\boldsymbol \theta}) = \langle -\ln p_{\boldsymbol \theta}({\bf v}) \rangle_{{\bf v} \sim p_{\mathcal D}} = \langle {\mathcal E}_{\boldsymbol \theta}({\bf v})  \rangle_{{\bf v} \sim p_{\mathcal D}} + \ln Z_{\boldsymbol \theta}
\end{equation}
where $\langle \cdot \rangle_{{\bf v} \sim p_{\mathcal D}} $ is the expectation value subject to data distribution $p_{\mathcal D}$. Due to the second term $\ln Z_{\boldsymbol \theta}$, the exact result of the loss function ${\mathcal L}({\boldsymbol \theta})$ can not be obtained unless for simple small systems. However, the gradient 
 \begin{equation}
 \label{eq:dL}
\nabla_{\boldsymbol \theta}{\mathcal L}({\boldsymbol \theta}) = \langle \nabla_{\boldsymbol \theta} {\mathcal E}_{\boldsymbol \theta}({\bf v})  \rangle_{{\bf v} \sim p_{\mathcal D}} - \langle \nabla_{\boldsymbol \theta} {\mathcal E}_{\boldsymbol \theta}({\bf v})  \rangle_{{\bf v} \sim p_{\boldsymbol \theta}}
\end{equation}
can be sampled in the same spirit that the thermal energy in statistical physics--the derivative of the partition function with respect to (inverse) temperature, can be sampled.  The resulting gradient descent algorithm updates parameters from step $t$ to step $t+1$ with learning rate $\eta$ as
 \begin{equation}
{\boldsymbol \theta}_{t+1} ={\boldsymbol \theta}_{t} 
 - \eta \nabla_{\boldsymbol \theta}{\mathcal L}({\boldsymbol \theta}_t).
\end{equation}
Th general gradient in this model
\begin{equation*}
 \nabla_{\boldsymbol \theta} {\mathcal E}_{\boldsymbol \theta}({\bf v}) = 
 \nabla_{\boldsymbol \theta} \left[
 -{\bf b}^T {\bf v} - \sum\limits_{i=1}^{n_h} \ln \left( 1+ e^{ {\bf w}_i^T {\bf v} + c_i }  \right) \right],
\end{equation*}
has components
\begin{equation}
\left\{
\begin{aligned}
\frac{\partial {\mathcal E}_{\boldsymbol \theta}({\bf v}) }{\partial W_{ij}} &= - \frac{ v_j e^{{\bf w}_i^T {\bf v} + c_i}}{1+ e^{{\bf w}_i^T {\bf v} + c_i} }  =-v_j\sigma( \alpha_i({\bf v}) ) \\
\frac{\partial {\mathcal E}_{\boldsymbol \theta}({\bf v}) }{\partial c_i} &= - \frac{e^{{\bf w}_i^T {\bf v} + c_i}}{1+e^{{\bf w}_i^T {\bf v} + c_i} } =-\sigma(\alpha_i({\bf v})) \\
 \frac{\partial {\mathcal E}_{\boldsymbol \theta}({\bf v}) }{\partial b_j} &= - v_j
 \end{aligned}
 \right.,
\end{equation}
where $\sigma(x) = 1/(1+e^{-x})$ is the sigmoid function.

The first term (data term) in Eq.~(\ref{eq:dL}) can be directly calculated by drawing ${\bf v}$ from dataset. The second term (model term) requires sampling ${\bf v}$ from $p_{\boldsymbol \theta}({\bf v})$. In the absence of this absolute probability density function one can generate valid ${\bf v}$'s from equilibrium Markov chains ${\bf h} \to {\bf v} \to {\bf h} \to {\bf v} \to \cdots$  according to conditional probabilities $ p_{\boldsymbol \theta}({\bf v} | {\bf h}) = e^{-E_{\boldsymbol \theta}({\bf v}, {\bf h})} / e^{-{\mathcal F}_{\boldsymbol \theta}({\bf h})}$ and $ p_{\boldsymbol \theta}({\bf h} | {\bf v}) =e^{-E_{\boldsymbol \theta}({\bf v}, {\bf h})} / e^{-{\mathcal E}_{\boldsymbol \theta}({\bf v})}$. Independence between individual hidden(visible) units are granted due to the lack of intralayer connections in RBMs, which leads to factorized conditional probabilities  $p_{\boldsymbol \theta}({\bf v} | {\bf h}) = \prod_j p_{\boldsymbol \theta}(v_j | {\bf h})$ and $p_{\boldsymbol \theta}({\bf h} | {\bf v}) = \prod_i p_{\boldsymbol \theta}(h_i | {\bf v})$. Further derivations show that the hidden unit follows a binomial distribution
\begin{align}
 p_{\boldsymbol \theta}(h_i | {\bf v}) =\frac{  e^{\alpha_i({\bf v}) h_i}}{1+e^{\alpha_i({\bf v})} }.
\end{align}
Namely, $p_{\boldsymbol \theta}(h_i=1 | {\bf v}) = \sigma(\alpha_i({\bf v}))$ and $p_{\boldsymbol \theta}(h_i=0 | {\bf v}) = 1-\sigma(\alpha_i({\bf v}))$.
The visible unit is exponentially distributed with probability density
\begin{align}
\begin{aligned}
 p_{\boldsymbol \theta}(v_j | {\bf h})  &= \frac{\beta_j({\bf h}) e^{\beta_j({\bf h}) v_j}}{e^{\beta_j({\bf h})} - e^{-\beta_j({\bf h})}}.
\end{aligned}
\end{align}
When the probability density function (pdf) and the cumulative distribution function (cdf) are available,  the standard cdf inversion method can be used to sample a random variable~\cite{andrieu2003}.
For the pdf $f(x) = \frac{k e^{kx}}{e^k - e^{-k}} $ ($-1\le x \le 1$),  the cdf is $F(x) = \int_{-1}^x f(x') dx' =\frac{ e^{kx} - e^{-k}}{e^k - e^{-k}} $. One can sample a uniformly distributed random number $u = F(x) \in [0,1]$ and find the exponentially distributed $x$ by $x = F^{-1}(u) = \frac{1}{k} \ln [(e^k - e^{-k})u + e^{-k}]$. Because $v_j$'s in our study are $x,y$-components of spins with unit length, an extra normalization $x_j = x_j/\sqrt{x_j^2 +y_j^2}$ and $y_j = y_j/\sqrt{x_j^2 +y_j^2}$ is added after every ${\bf h} \to {\bf v}$ generation step.

\subsection{Real-valued restricted Boltzmann machines with nonlinear activation (RBM-CosSin)}
\begin{figure}
\includegraphics[width=1\linewidth]{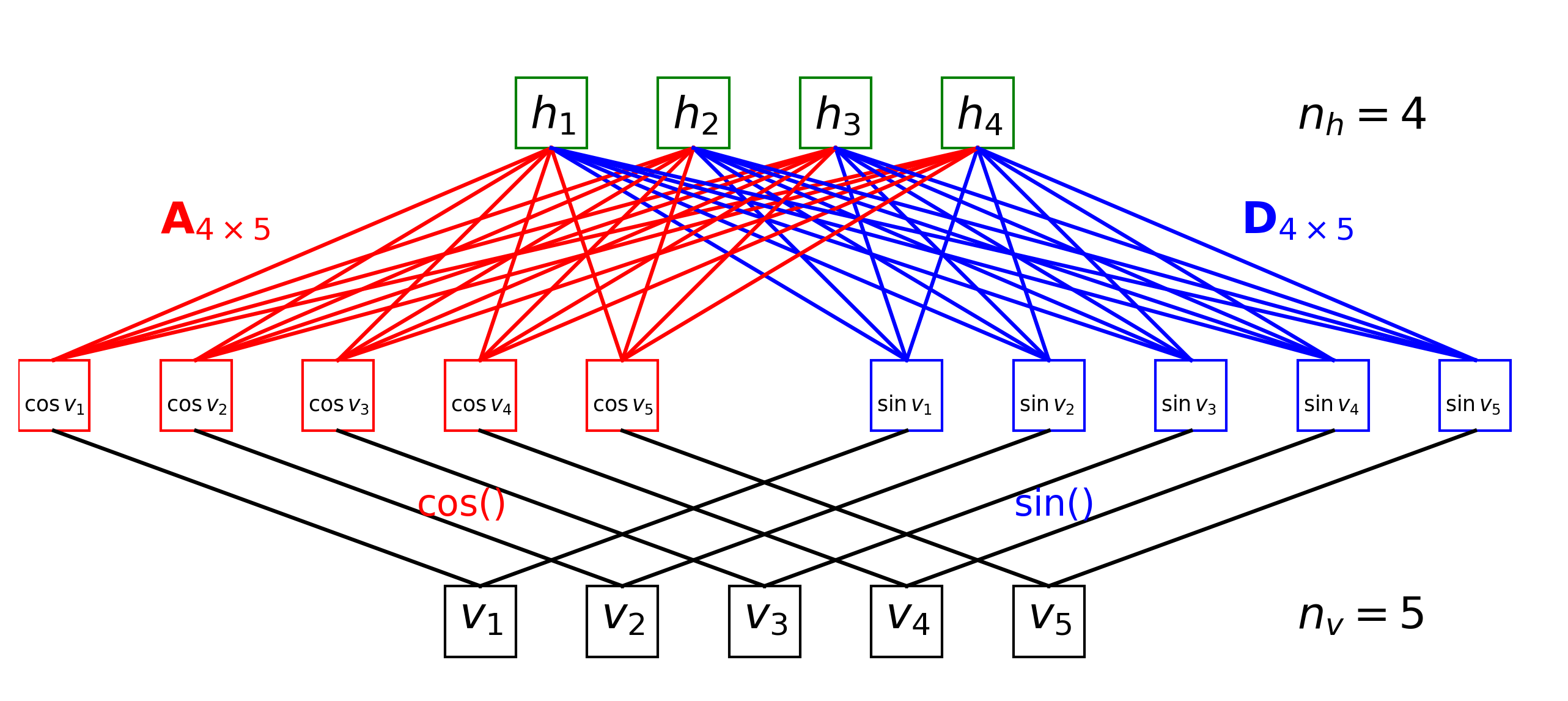}%
\caption{\label{fig:rbmcs} Graphical representation of the RBM-CosSin with $n_v=5$ visible units and $n_h=4$ hidden units. Nonlinear $\cos$/$\sin$ activation is applied to visible units ${\bf v}$ before connected with the hidden layer.}
\end{figure}

In this type of real-valued RBMs (termed as ``RBM-CosSin''), the hidden units still have binary values $\{0,1\}$. The visible units are chosen as the angles of spins i.e.  $v_j = \varphi_j \in[0,2\pi]$ ($j=1,2,\cdots,n_v$) and $n_v = N$  (Fig.~\ref{fig:rbmcs}). Motivated by the requirement that the energy should be periodic with respect to angles, the total energy, after neglecting unimportant bias terms, is defined as
\begin{equation}
\begin{aligned}
& E_{\boldsymbol{\theta}}({\bf v}, {\bf h}) =   - {\bf h}^T{\bf A} \cos {\bf v} - {\bf h}^T{\bf D} \sin {\bf v} \\
&=  -\sum\limits_{i=1}^{n_h} \sum\limits_{j=1}^{n_v} A_{ij} h_i \cos v_j  -\sum\limits_{i=1}^{n_h} \sum\limits_{j=1}^{n_v} D_{ij} h_i \sin v_j \\
& = -  {\bf h}^T_{1\times n_h} 
\begin{bmatrix}
    {\bf A} & {\bf D}
\end{bmatrix}_{n_h\times 2n_v}
\begin{bmatrix}
\cos {\bf v}  \\
 \sin {\bf v} 
\end{bmatrix}_{2n_v \times 1},
\end{aligned}
\end{equation}
for which ${\boldsymbol \theta} = \{{\bf A}, {\bf D}\}$. The decomposition of ${\bf A}, {\bf D}$ matrix into row vectors (filters) or column vectors (inverse filters) is similar as in the above case of ${\bf W}$ matrix.  Functions applied upon vectors should be understood as  element-wise operations, e.g. $\cos {\bf v} = [\cos v_1, \cos v_2, \cdots, \cos v_{n_v}]^T$.
It can be seen that this RBM-CosSin is related to  RBM-xy by mapping $\begin{bmatrix}
    {\bf A} & {\bf D} 
\end{bmatrix}$ onto ${\bf W}$. For systems of $N$ spins, both RBM-xy and RBM-CosSin have the same number ($n_h \times 2N$) of matrix parameters. However, RBM-CosSin does not need the ad hoc normalization process used in RBM-xy.

Let  $\gamma_i({\bf v}) = {\bf a}_i^T \cos {\bf v} + {\bf d}_i^T \sin {\bf v} $, $\xi_j({\bf h}) =   {\bf h}^T {\bf a}_{:,j}$ and  $\zeta_j({\bf h}) =   {\bf h}^T {\bf d}_{:,j}$. The total energy can be rewritten as 
\begin{equation}
\begin{aligned}
E_{\boldsymbol{\theta}}({\bf v}, {\bf h}) & = - {\bf h}^T {\boldsymbol \gamma}({\bf v})  = - {\boldsymbol \xi}^T({\bf h}) \cos {\bf v}  - {\boldsymbol \zeta}^T({\bf h}) \sin {\bf v} 
\end{aligned}
\end{equation}
where the vector ${\boldsymbol \gamma}=[\gamma_1, \gamma_2, \cdots, \gamma_{n_h}]^T$, etc. This notation helps us quickly see the factorization 
\begin{equation*}
\begin{aligned}
\sum_{{\bf h}} e^{-E_{\boldsymbol{\theta}}({\bf v}, {\bf h})} &= \sum_{{\bf h}} e^{{\bf h}^T {\boldsymbol \gamma}({\bf v})}   = \prod_{i=1}^{n_h} (1 + e^{   \gamma_i({\bf v})} )  
\end{aligned}
\end{equation*}
from which the visible energy can be derived
\begin{equation}
\begin{aligned}
 {\mathcal E}_{\boldsymbol \theta}({\bf v}) & = - \sum\limits_{i=1}^{n_h} \ln \left(1+e^{ {\bf a}_i^T \cos {\bf v} + {\bf d}_i^T \sin {\bf v} }\right) \\
&=  - \sum\limits_{i=1}^{n_h} \ln \left(1+e^{   \gamma_i({\bf v})}\right).
\end{aligned}
\end{equation}

The conditional probability to sample $h_i$ at fixed ${\bf v}$ is the similar binomial distribution as before
\begin{align}
\begin{aligned}
 p_{\boldsymbol \theta}(h_i=1 | {\bf v}) = \sigma(\gamma_i({\bf v})   )  .
\end{aligned}
\end{align}
The sampling of $v_j$ at fixed ${\bf h}$ follows the  von Mises distribution (rounded to the $[0,2\pi]$ interval)
\begin{align}
\begin{aligned}
p_{\boldsymbol \theta}(v_j | {\bf h})  &= \frac{e^{\xi_j({\bf h}) \cos v_j +\zeta_j({\bf h}) \sin v_j }}{ I(\kappa_j({\bf h}))  } 
= \frac{e^{ \kappa_j \cos(v_j - \mu_j)}}{I(\kappa_j)}
\end{aligned}
\end{align}
where $\kappa_j = \sqrt{\xi_j^2 + \zeta_j^2} \ge 0$ is the inverse dispersion and $\mu_j$ is the mean angle that can be solved from $\cos \mu_j = \xi_j / \kappa_j$ and $\sin \mu_j = \zeta_j / \kappa_j$.
The denominator is an integral $I(\kappa) = \int_0^{2\pi} e^{\kappa \cos(x - \mu)} dx$  proportional to the Bessel function. One can make use of Python's built-in von Mises distribution sampler {\verb |numpy.random.vonmises()|} without calculating $I(\kappa)$ explicitly.

The gradients required in RBM-CosSin training are
\begin{equation}
\left\{
\begin{aligned}
\frac{\partial {\mathcal E}_{\boldsymbol \theta}({\bf v}) }{\partial A_{ij}} &= - \frac{ \cos v_j e^{\gamma_i({\bf v}) }}{1+ e^{\gamma_i({\bf v}) } }  =-\cos v_j \sigma(\gamma_i({\bf v}) ) \\
\frac{\partial {\mathcal E}_{\boldsymbol \theta}({\bf v}) }{\partial D_{ij}} &= - \frac{ \sin v_j e^{\gamma_i({\bf v}) }}{1+ e^{\gamma_i({\bf v}) } }  =-\sin v_j \sigma(\gamma_i({\bf v}) ) 
 \end{aligned}
 \right..
\end{equation}

\section{Results and Discussion}
\label{sec:result}

\subsection{Training of RBMs}
At each temperature, we train RBMs (RBM-xy or RBM-CosSin) using the dataset ${\mathcal D} = \{{\bf v}^{(1)}, {\bf v}^{(2)}, \cdots, {\bf v}^{(20000)} \}$ consisting of raw XY configurations. The number of hidden units are $n_h=256, 512$ for $L=16$ and $n_h=1024$ for $L=32$. Model parameters are initialized with Glorot normal initialization~\cite{glorot2010} and then optimized with the stochastic gradient descent of learning rate $\eta = 0.001$ and batch size 128 for 1000 epochs. 
In the model term $\langle \nabla_{\boldsymbol \theta} {\mathcal E}_{\boldsymbol \theta}({\bf v})  \rangle_{{\bf v} \sim p_{\boldsymbol \theta}}$ calculation stage of gradient descent, 1-step contrastive divergence  (CD-1) Gibbs sampling was used to sample ${\bf v}$ from $p_{\boldsymbol \theta}({\bf v})$~\cite{hinton2012}. In particular, the Markov chain starts from XY configurations in the dataset ${\bf v} \in {\mathcal D}$ and runs forward by one step ${\bf v} \to {\bf h} \to {\bf v}'$. All learning tasks are performed on Nvidia GeForce RTX 4090 GPU cards, which typically took less than ten seconds per epoch. 

\begin{figure}
\includegraphics[width=0.9\linewidth]{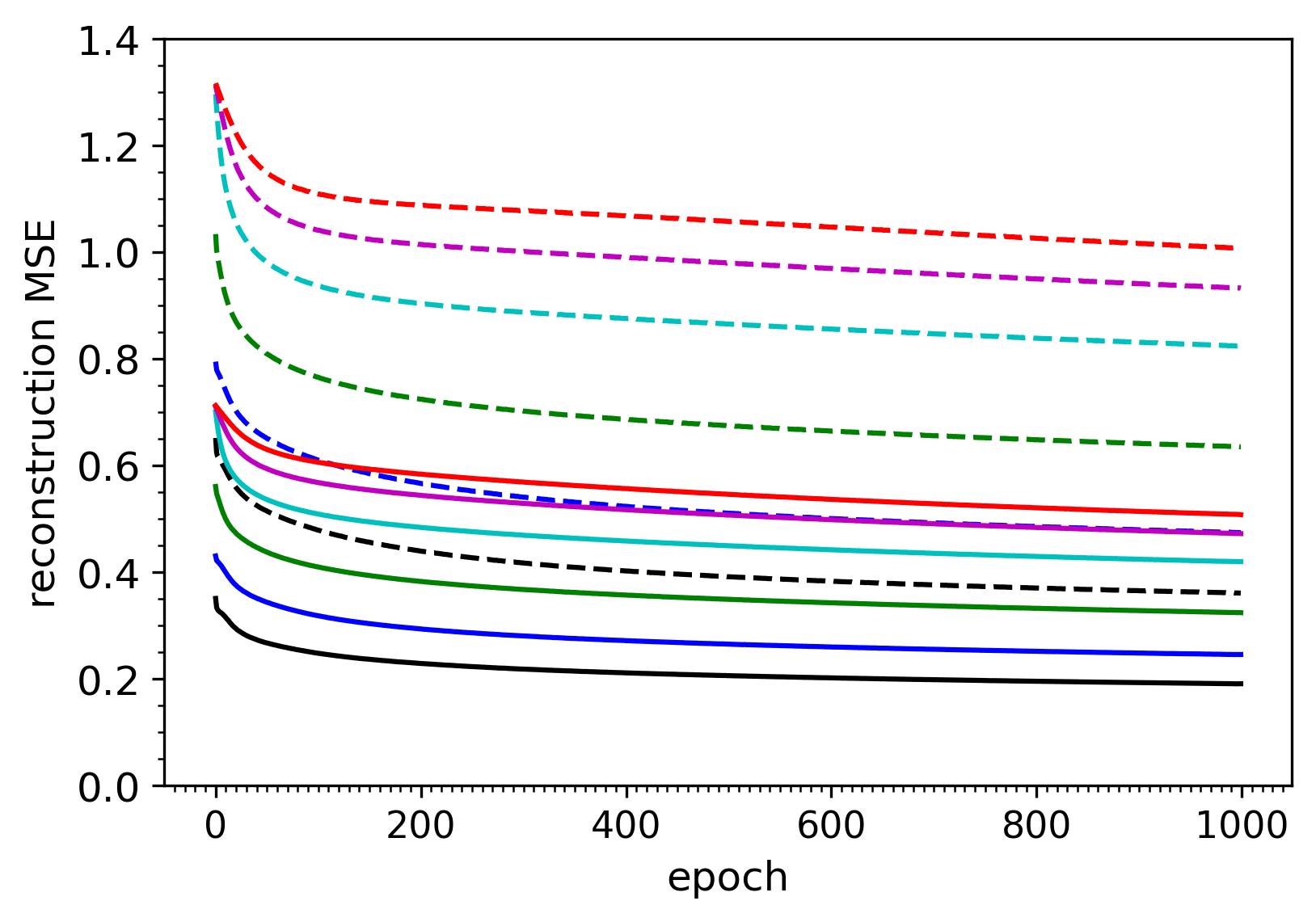}
\caption{\label{fig:mse} Example reconstruction mean-squared error (MSE) as a function of training epoch for RBM-xy (solid) RBM-CosSin (dashed) at temperatures $T=0.6,0.8,1.0,1.2,1.4,1.5,1.6$ (from bottom to top). $L=32$ and $n_h=1024$.  }
\end{figure}

We monitor the reconstruction mean-squared error (MSE) during training, which compares the average difference between original XY configurations ${\bf v}\in {\mathcal D}$ and RBM reconstructed configurations ${\bf v} \to {\bf h} \to {\bf v}'$. For RBM-xy,
\begin{align}
    {\rm MSE} = \left\langle \frac{1}{n_v} \sum_{j=1}^{n_v} (v_j' -v_j)^2 \right\rangle_{{\bf v}\sim p_{\mathcal D}}
\end{align}
For RBM-CosSin in which $v_j$ are angles,
\begin{align}
    {\rm MSE} = \left\langle \frac{1}{n_v} \sum_{j=1}^{n_v} \left[(\cos v_j' - \cos v_j)^2 +(\sin v_j' - \sin v_j)^2 \right]\right\rangle_{{\bf v}\sim p_{\mathcal D}}
\end{align}
Up to 1000 epoches, MSE drops monotonically until saturation with RBM-CosSin MSE being higher (Fig.~\ref{fig:mse}). Although reconstruction MSE is a convenient metric to evaluate the performance of  RBMs, it is not  the actual loss function being optimized. An apparently lower MSE of RBM-xy is due to overfitting, because further analysis below shows that RBM-CosSin generates configurations that better follow  the true Boltzmann distribution. 

\subsection{RBM generated configurations}
\begin{figure}
\includegraphics[width=0.85\linewidth]{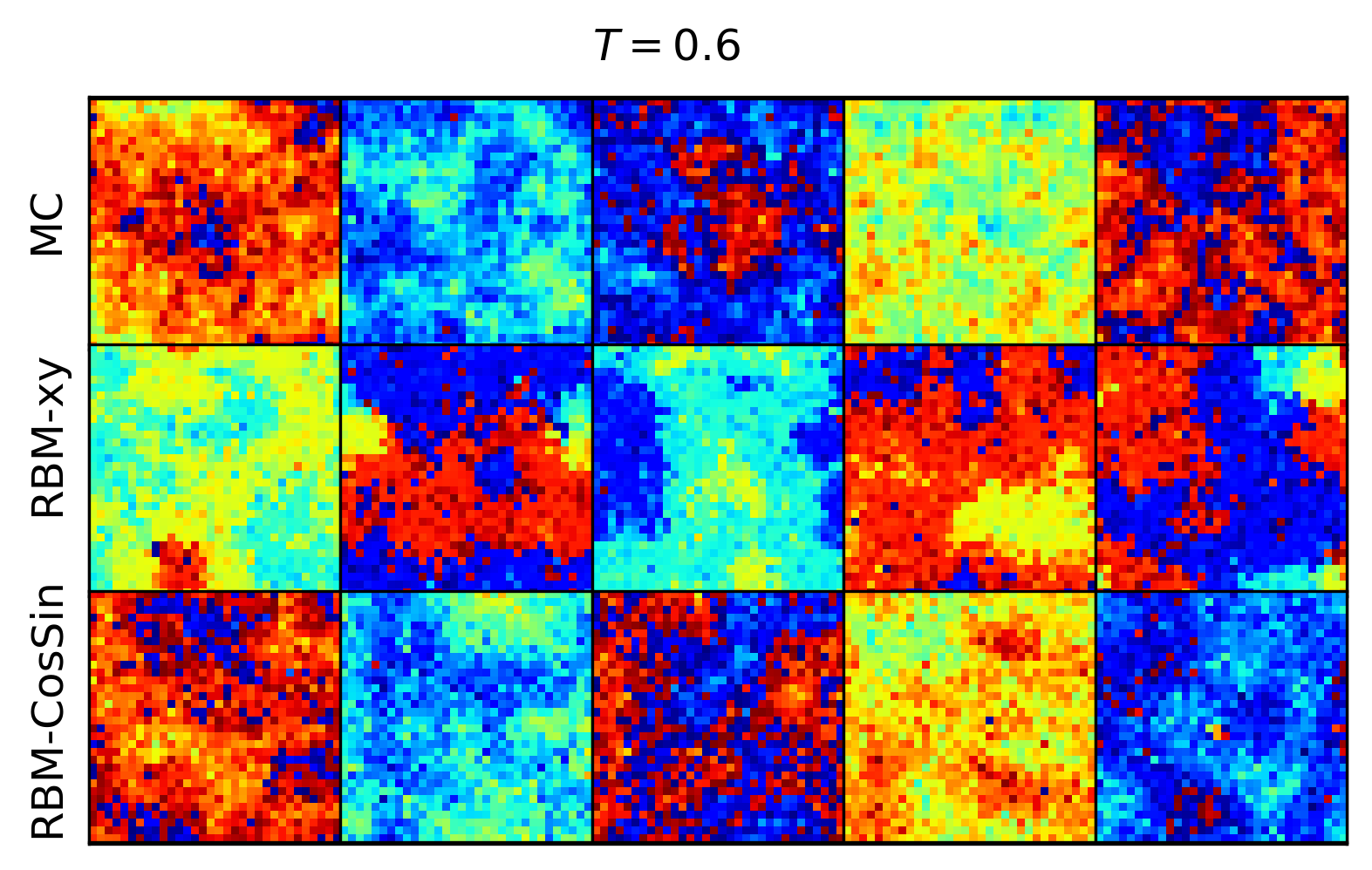}
\includegraphics[width=0.85\linewidth]{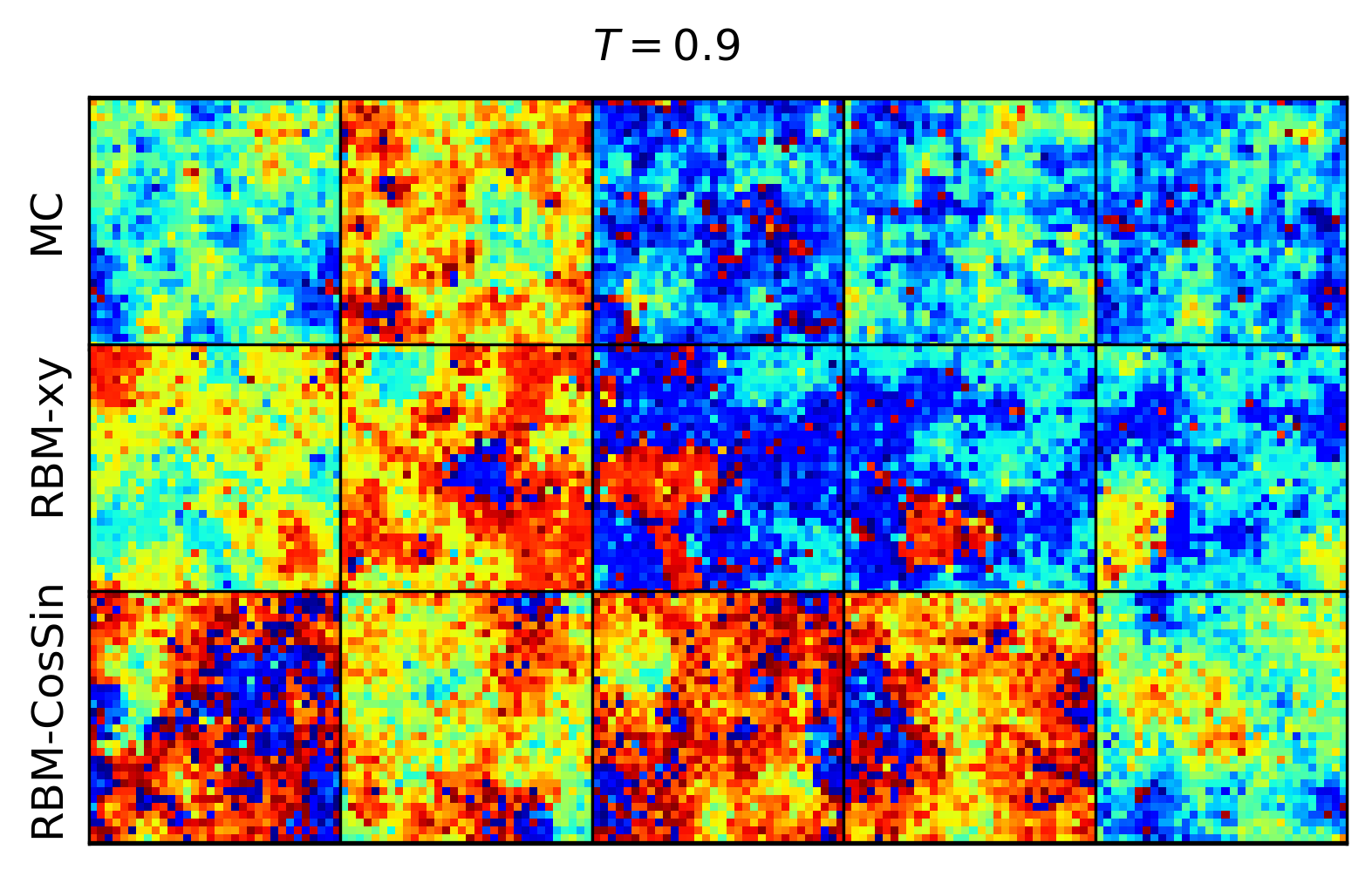}
\includegraphics[width=0.85\linewidth]{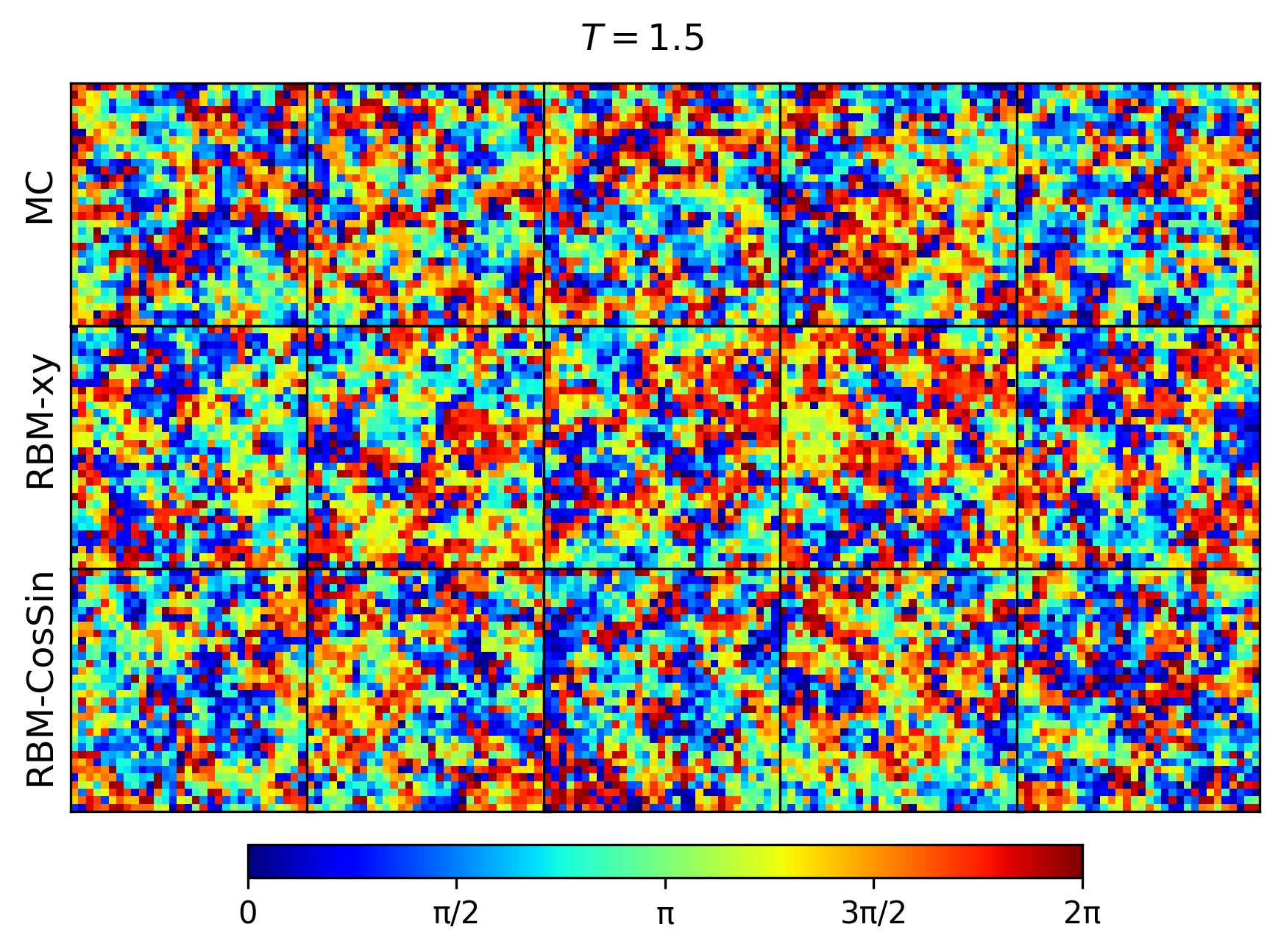}
\caption{\label{fig:config} Example XY configurations of $32\times32$ spins, in which spin angles $\varphi_i\in [0,2\pi]$ are color-coded, generated by Monte Carlo simulation (MC), RBM-xy and RBM-CosSin at low, medium and high temperatures.   }
\end{figure}
After our RBMs are trained, we start from random hidden units $h_i^{0} \sim B(1,0.5)$ (binomial distribution of size 1 and probability 0.5) and run the Markov chain ${\bf h}^{0} \to  {\bf v}^{0} \to {\bf h}^{1} \to {\bf v}^{1} \cdots$ for 100 steps. At the end, 20000 visible states are collected as generated configurations at each temperature. At high $T$, both RBM-xy and RBM-CosSin generate configurations that are very close to   true XY configurations as indicated by the heatmaps of spin angles (Fig.~\ref{fig:config}). At low $T$, RBM-CosSin generated configurations are more genuine, while RBM-xy generated configurations contain clusters of homogeneous spins that are in discrepancy with intracluster spin fluctuations of real XY configurations.   
Note that the colorbar scheme in our plots is chosen in order to quantitatively distinguish all possible values of angles within $[0,2\pi]$. It should be kept in mind that the high contrast between red ($2\pi$) and blue ($0$) colors does not imply a large difference in angles due to periodicity.

\begin{figure}
\includegraphics[width=0.85\linewidth]{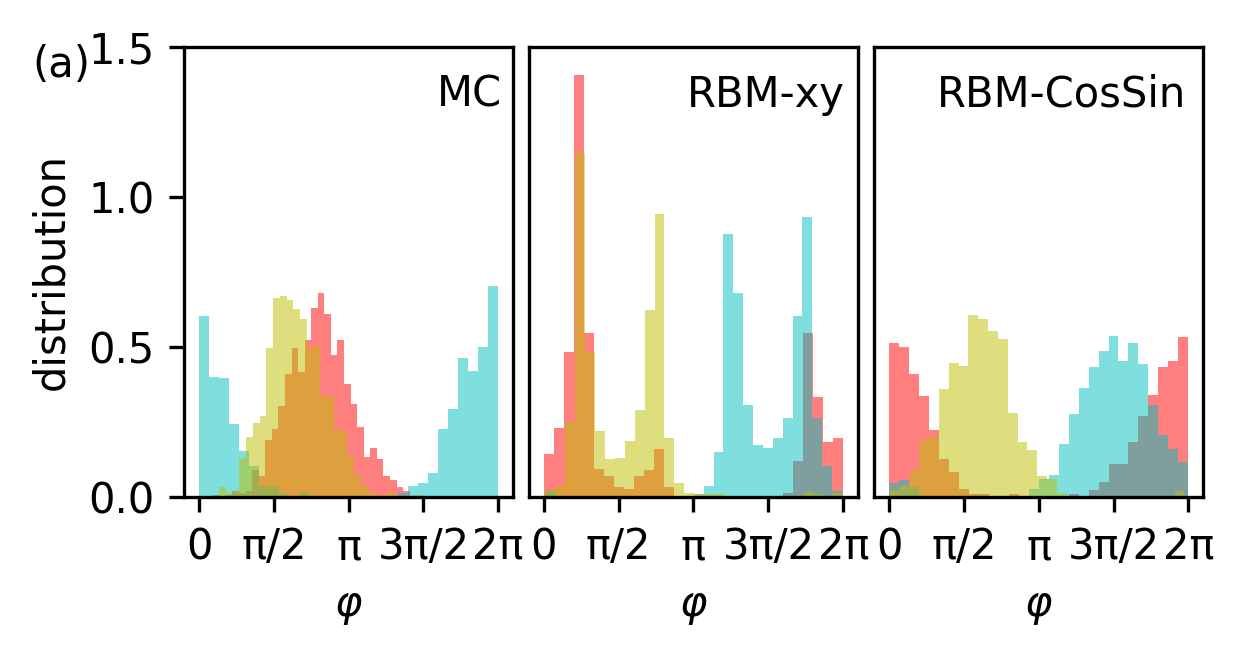}
\includegraphics[width=0.85\linewidth]{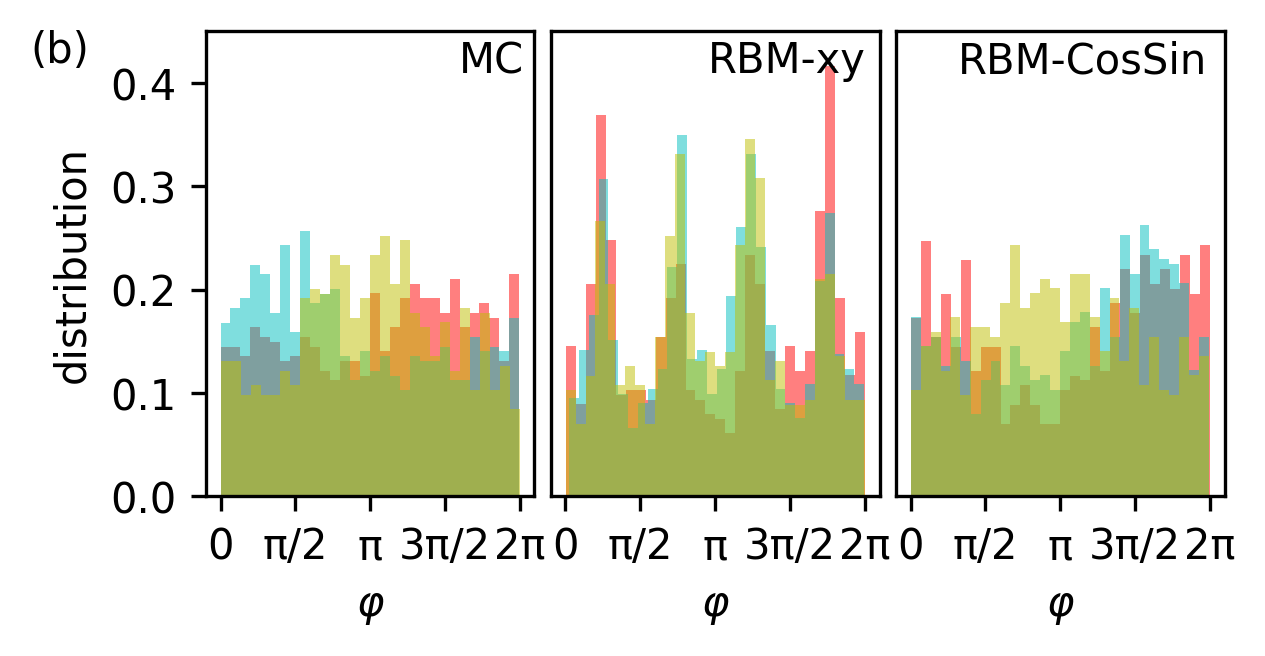}
\caption{\label{fig:phi} Distributions of spin angle $\varphi\in [0,2\pi]$ in three sample XY configurations ($L=32$) generated by Monte Carlo simulation (MC), RBM-xy and RBM-CosSin at (a) $T=0.6$  and (b) $T=1.5$.     }
\end{figure}

To quantify generated spin configurations, we study the histogram distribution of $\varphi$. Below $T_c$, angles are narrowly distributed around a mean value, which should be uniformly distributed within $[0,2\pi]$ due to rotational invariance of the XY Hamiltonian (Fig.~\ref{fig:phi}a). RBM-CosSin can well capture this unimodal distribution using its von Mises distribution. In contrast, the distribution of angles in RBM-xy configurations is bimodal (sometimes trimodal). Although a bimodal distribution can give rise to the same mean as a unimodal one, the dispersion is different. This explains the abnormal homogeneous spin clusters of RBM-xy configurations in Fig.~\ref{fig:config} because each peak of the bimodal distribution is narrower.

Above $T_c$, XY angles are uniformly distributed (with certain noises), which is well approximated by the von Mises distribution with small $\kappa_j$ in RBM-CosSin (Fig.~\ref{fig:phi}b). The distribution in RBM-xy configurations is multimodal with four characteristic peaks, although the coverage of the entire $[0,2\pi]$ range is relatively more uniform than the low temperature case. The inaccuracy of spin angles generated by RBM-xy is due to its  monotonically increasing/decreasing exponential distribution function of $x,y$-components, which allocates maximum probability density at either $1$ or $-1$. After normalized and converted back to angles, the resulting distribution of $\varphi$ cannot be unimodal or flat. 

\subsection{Thermodynamics learned by RBMs}
\begin{figure}
\includegraphics[width=0.45\linewidth]{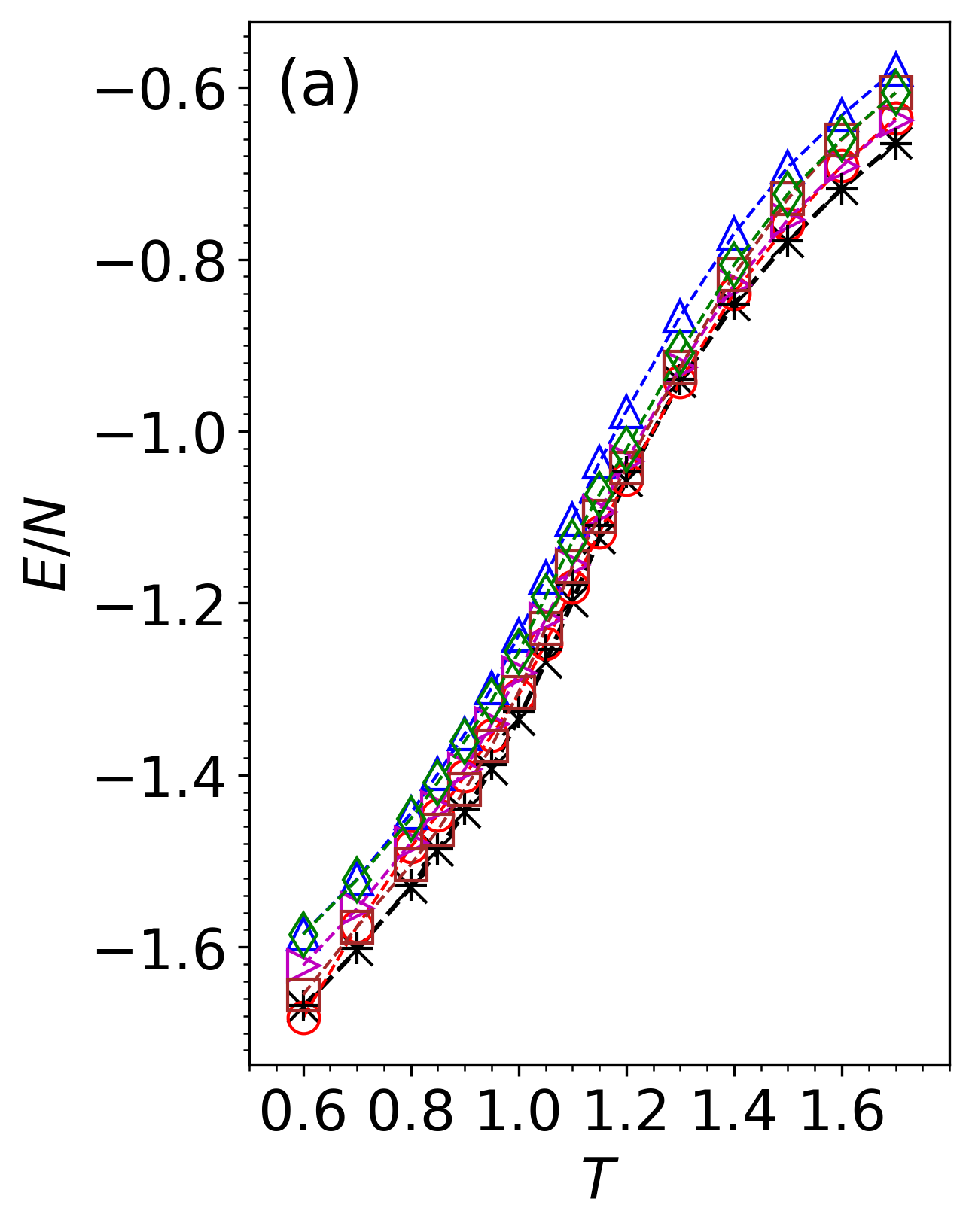}%
\includegraphics[width=0.43\linewidth]{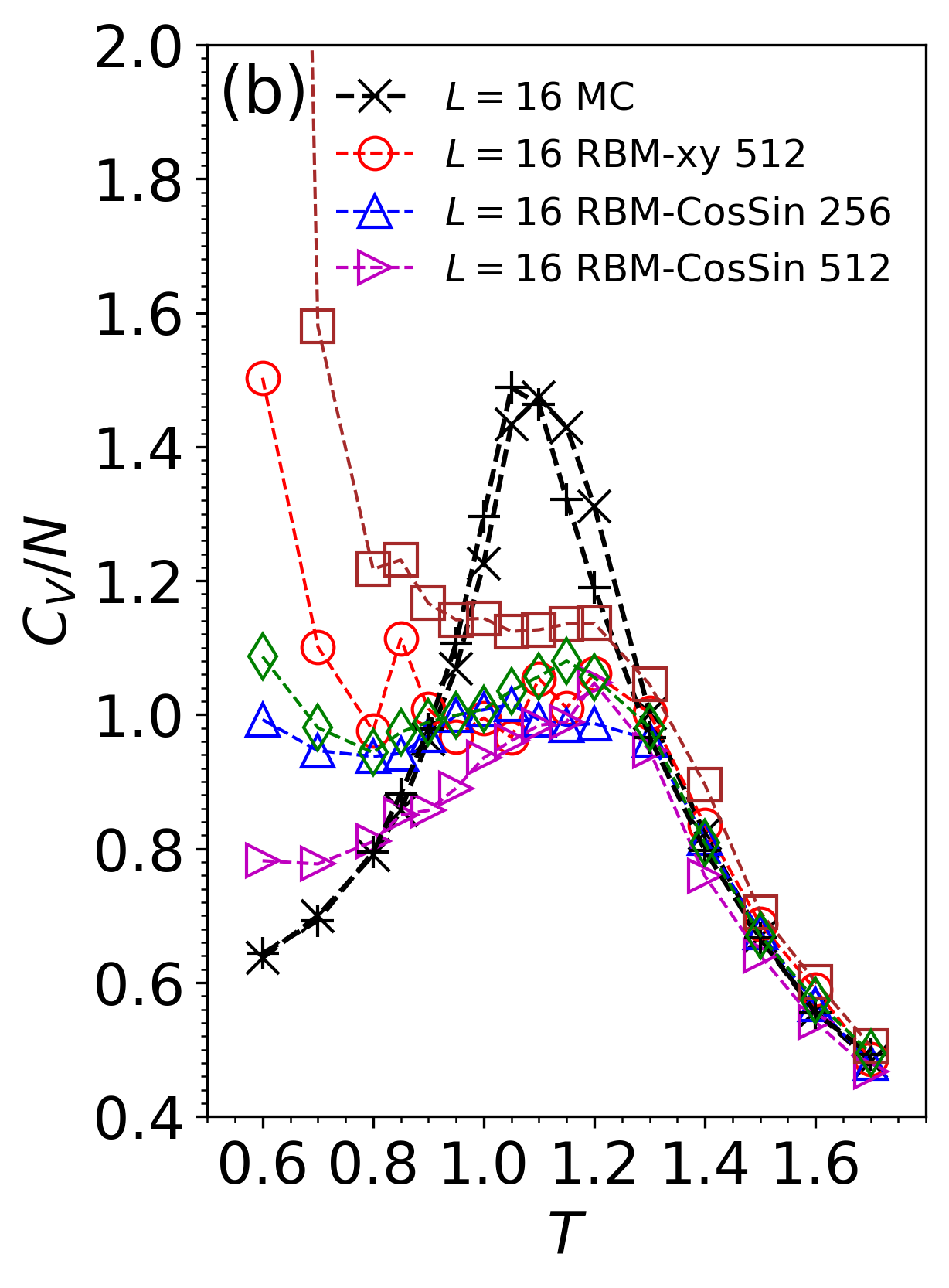}
\includegraphics[width=0.45\linewidth]{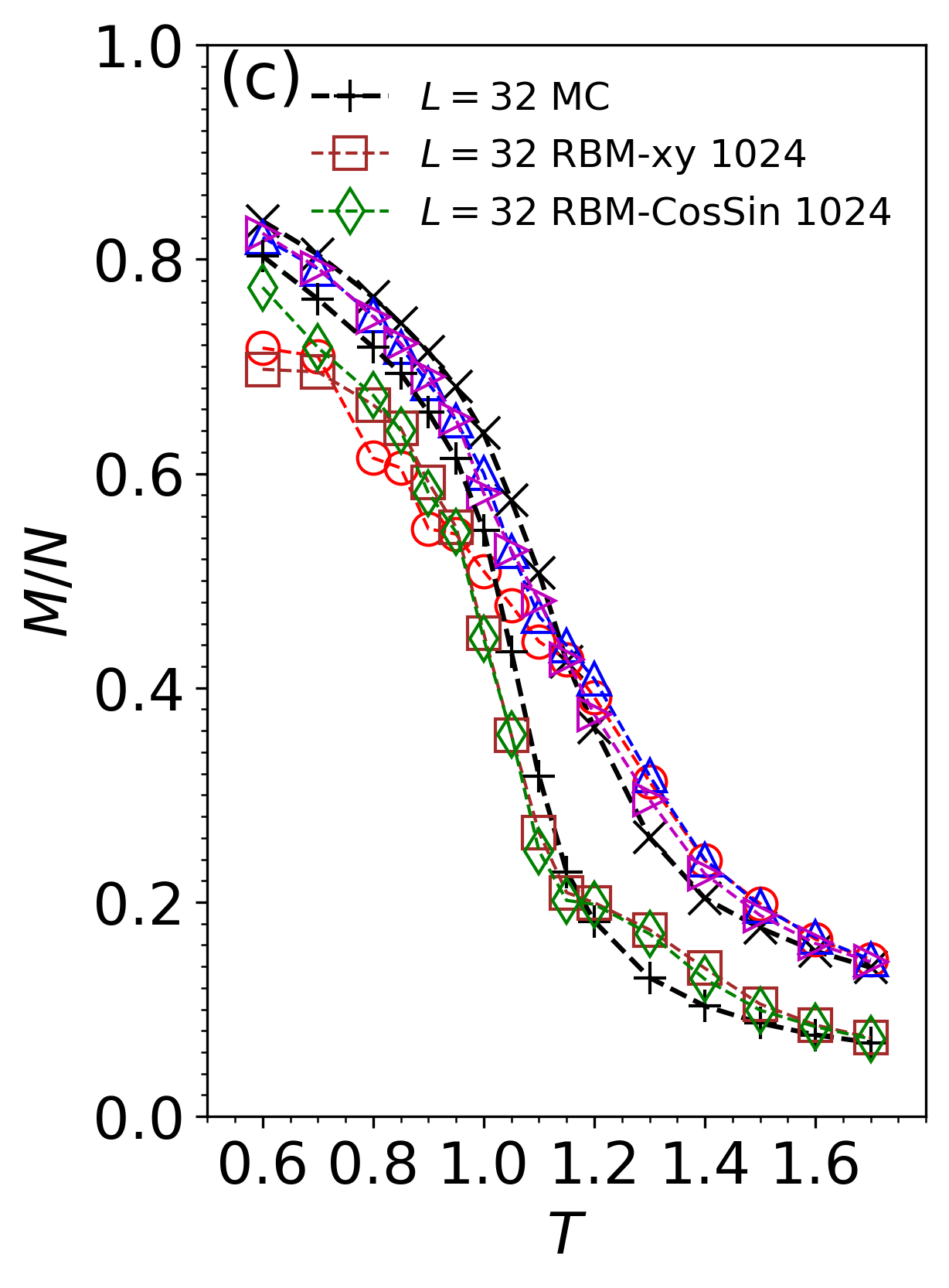}
\includegraphics[width=0.43\linewidth]{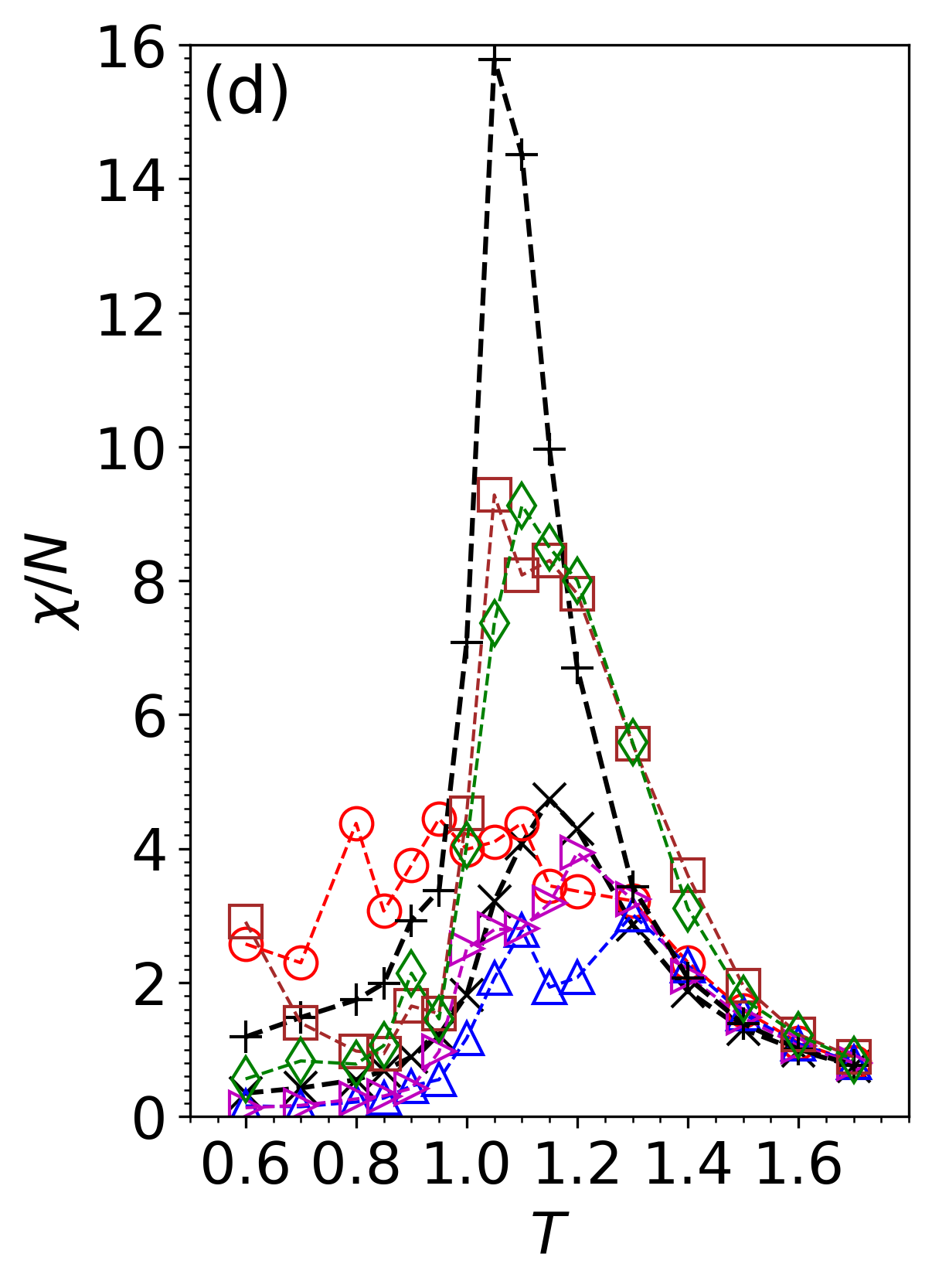}
\caption{\label{fig:thermo} Average energy (a), heat capacity (b), magnetization (c) and susceptibility (d) per spin of 2D XY configurations generated by RBM-xy and RBM-CosSin with $n_h=256,512,1024$ hidden units over temperature range $T=0.6$-$1.7$ for systems of linear dimension  $L=16, 32$. Results from Monte Carlo simulation (MC) are also shown for comparison.}
\end{figure}

We then calculate the thermodynamic properties of XY configurations generated by RBMs and compare them with Monte Carlo results. All RBMs can capture internal energy fairly well. For the same model, increasing hidden units number $n_h$ improves the accuracy (Fig.~\ref{fig:thermo}a). The performance of RBM-XY can even slightly surpass that of RBM-CosSin when it comes to energy calculations alone. However, in combination with other metrics, we shall see that the shortcomings of RBM-XY become more pronounced at low temperatures. 

RBMs can only achieve accurate heat capacity measurements for $T \gtrsim 1.3$ (Fig.~\ref{fig:thermo}b). The largest deviation from Monte Carlo results occurs at the lowest temperature $T=0.6$ in this study, where RBM-xy estimations begin to blow up. RBM-CosSin partially suppresses this abnormal divergence and is able to preserve a smooth peak of $C_V$ around $T_p$. This suggests that capturing energy fluctuations for low-temperature XY states is considerably more challenging. The situation in RBM learning of the Ising model is different, as low-temperature ferromagnetic phases can be described equally well~\cite{torlai2016,morningstar2018}. Only near the critical temperature, RBM heat capacity differs from Monte Carlo results significantly. 

As mentioned above, in the thermodynamic limit, 2D XY magnetization goes to zero. Apparently non-zero magnetization is observed in finite systems, whose value decreases with system size (Fig.~\ref{fig:thermo}c). As the temperature decreases, the deviation in RBM-xy magnetization becomes increasingly significant, while the estimations from RBM-CosSin remain close to the Monte Carlo results across the entire temperature range. Similarly, the finite-size susceptibility (with non-zero magnetization) is much more noisy in the case of RBM-xy, especially around the peak and the low temperature regime (Fig.~\ref{fig:thermo}d). Conversely, RBM-CosSin is able to locate   the peak of $\chi$ semi-quantitatively. 

\begin{figure}
\includegraphics[width=0.6\linewidth]{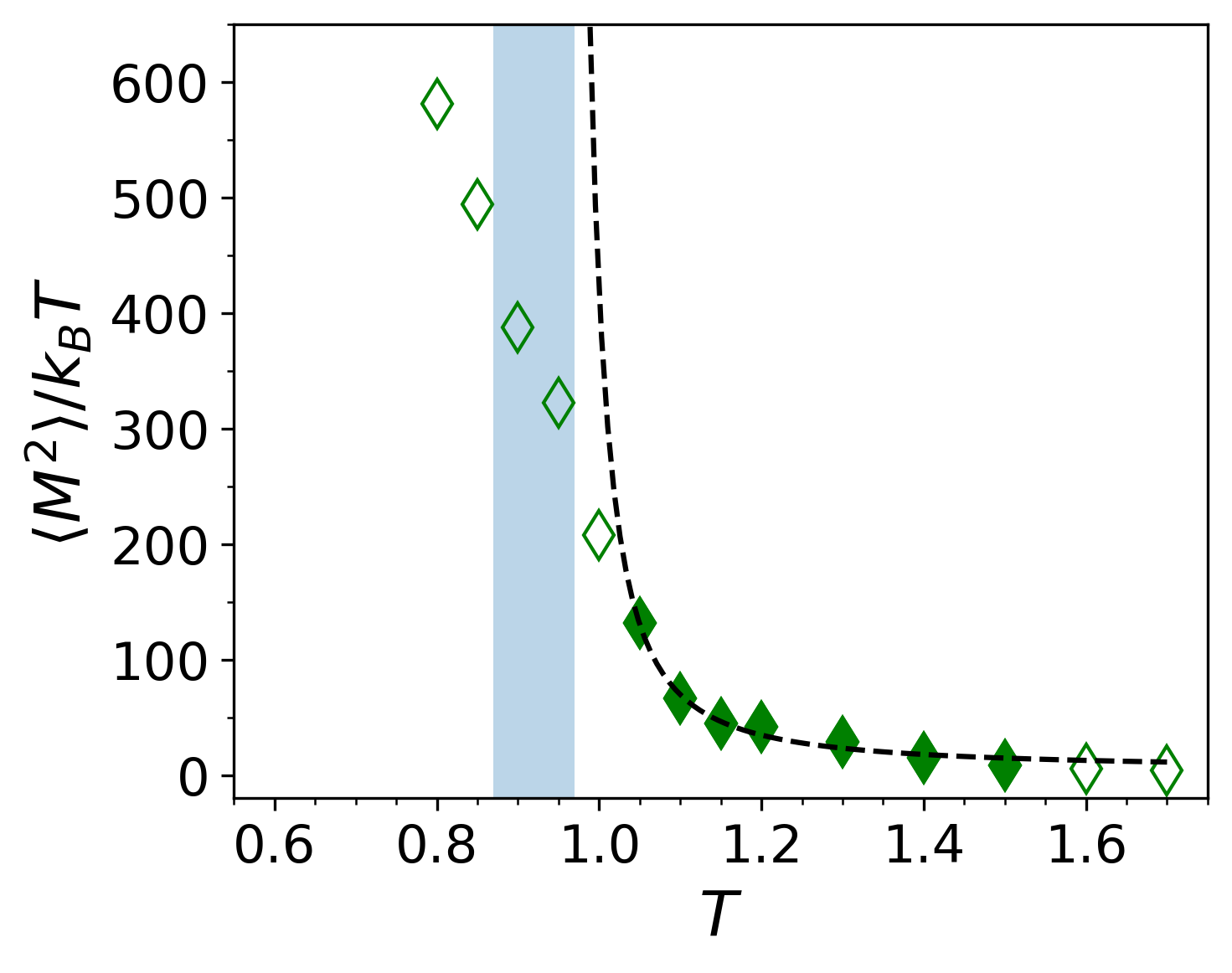}
\caption{\label{fig:m2} Mean-squared magnetization $\frac{\langle M^2\rangle}{k_BT}$ (as the approximation to $\chi$) for $n_h=1024$ RBM-CosSin generated XY configurations of size $L=32$. The dashed line is the fitting of data within $1.05\le T\le 1.5$ (solid symbols) according to KT theory $\chi(T) = a_\chi e^{b_\chi/\sqrt{T/T_c -1}}$ with $a_\chi=2(2)$,  $b_\chi=1.5(7)$ and $T_c=0.92(5)$ (highlighted by the transparent rectangle). }
\end{figure}

Before more reliable methods were devised, the divergence of susceptibility (assuming zero magnetization) was used to provide the first estimation to the KT transition temperature $T_c$~\cite{tobochnik1979,gupta1988}. One can replace the finite-size $\chi$, which exhibits a peak as shown above, with the mean-squared magnetization $\frac{\langle M^2\rangle}{k_BT}$ to investigate this behavior. We test this idea with our RBM-CosSin generated XY configurations of the $L=32$ system. Mean-squared magnetization data points above and close to $T_c$ are used in the fitting based on the KT theory $\chi(T) = a_\chi e^{b_\chi/\sqrt{T/T_c -1}}$, which diverges at the estimated $T_c = 0.92(5)$ (Fig.~\ref{fig:m2}). This is another evidence that RBM-CosSin can learn the underlying Boltzmann distribution of XY configurations and even retain valuable information about the KT transition. 

\subsection{Topological order learned by RBMs}
\begin{figure}
\includegraphics[width=0.6\linewidth]{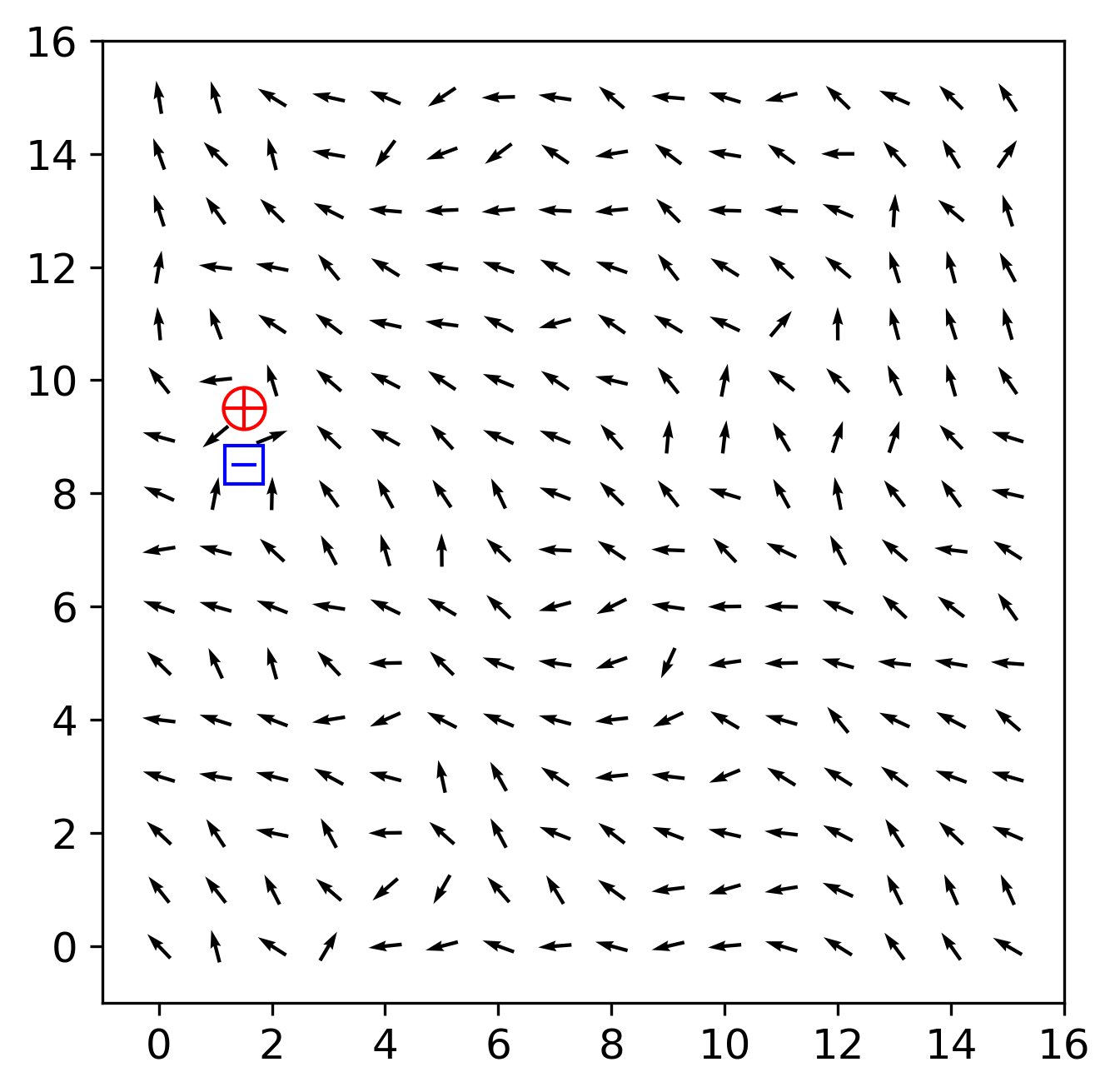}
\includegraphics[width=0.6\linewidth]{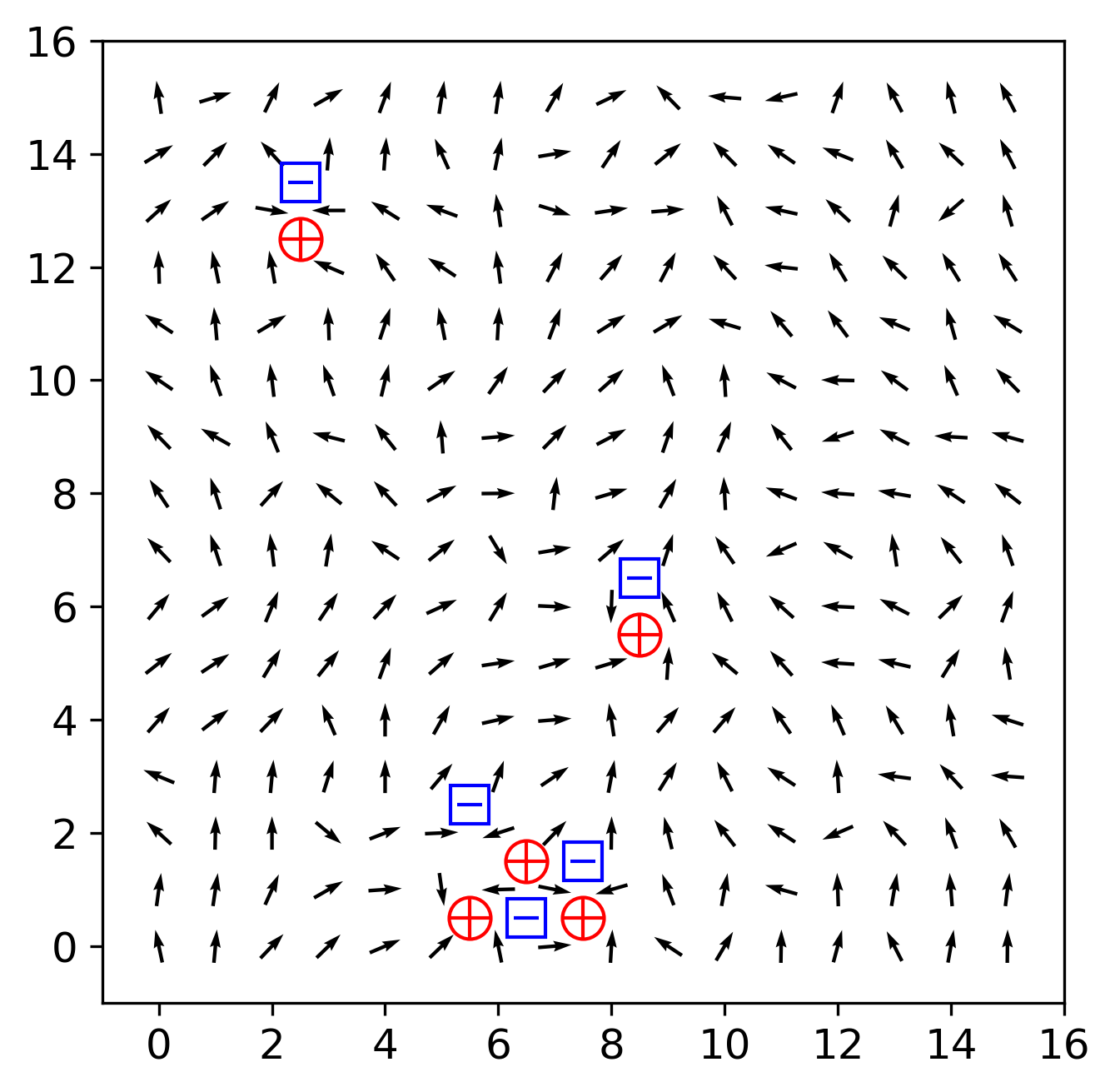}
\includegraphics[width=0.6\linewidth]{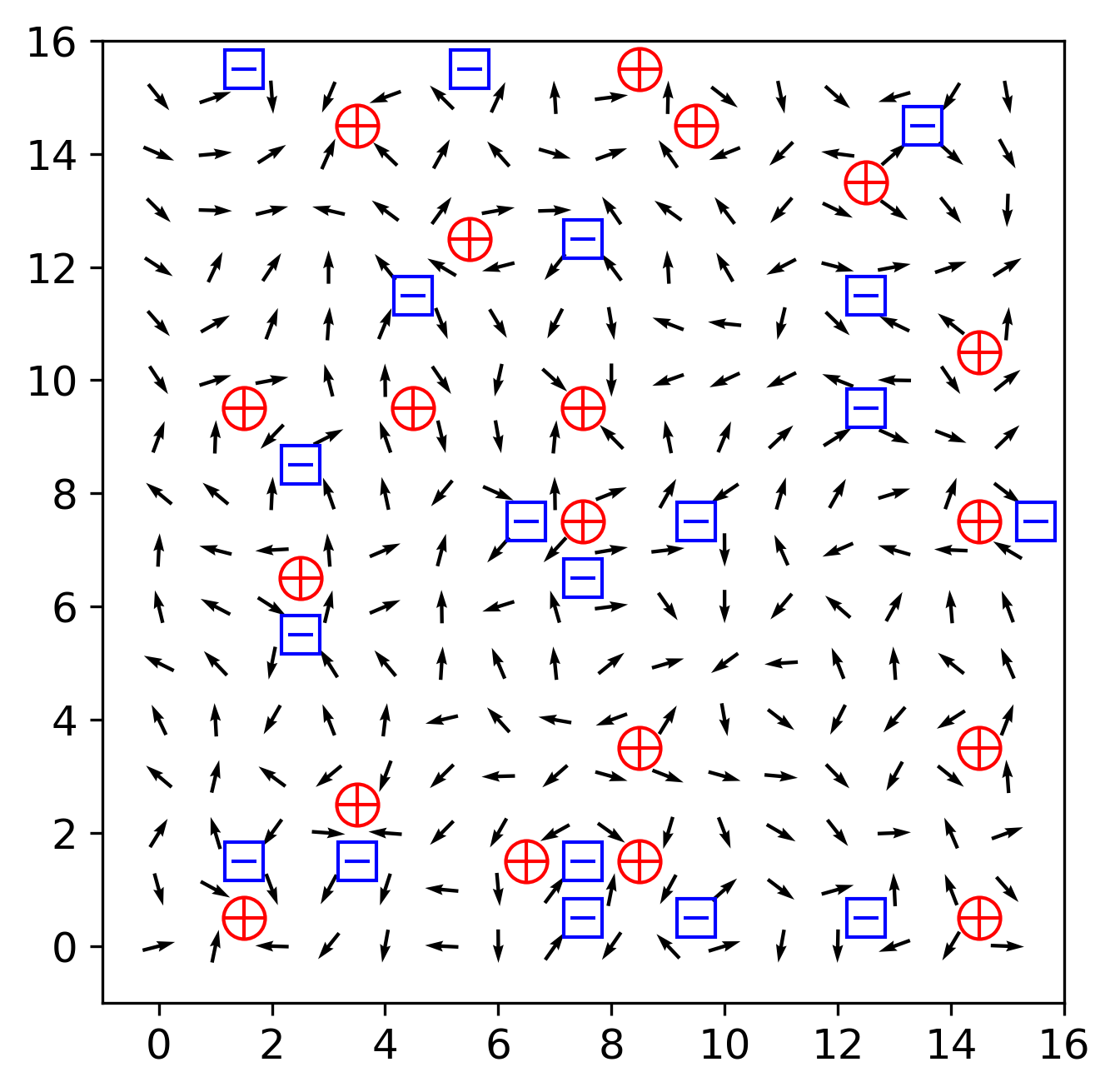}
\caption{\label{fig:vortex} Examples of vortices (plus-circle) and anti-vortices (minus-square) in XY configurations of $16\times 16$ spins (short arrows) generated by $n_h=512$ RBM-CosSin at $T=0.6,0.9,1.5$ (from top to bottom). }
\end{figure}
Besides thermodynamic properties, the most unique and characteristic feature of the 2D XY model is its topological order manifested through spin vortices because it is directly related to the KT transition. Roughly speaking, spin vectors rotate counter-clock-wisely (clock-wisely) by $2\pi$ around a vortex (anti-vortex) center when tracing around an enclosing loop $C$ in a counterclockwise direction. The quantitative measure of vorticity is the winding number $k$ (topological charge) defined as $\oint_C \nabla \varphi \cdot d\vec{l} = 2\pi k$, $k=\pm1, \pm 2,\cdots$. A vortex (anti-vortex) corresponds to the case of $k=+1$ ($-1$). On a square lattice, the integral of angle gradient is approximated by the sum of angle differences over the four neighboring spin pairs forming a plaquette. In particular, for a spin on the lattice site $(i,j)$, we calculate  $\Delta \varphi = [\varphi_{i,j+1} - \varphi_{i,j}] + [\varphi_{i+1,j+1} - \varphi_{i,j+1}]+[\varphi_{i+1,j} - \varphi_{i+1,j+1}]+[\varphi_{i,j} - \varphi_{i+1,j}]$ under periodic boundary conditions, where $[\cdot]$ means to round the angle difference to the interval $[-\pi, \pi]$ using a saw function~\cite{beach2018}. With this method, we detect all possible vortex/anti-vortex in each XY configuration and count the total number of vortices ($k_+$) and anti-vortices ($k_-$). Periodic boundary condition imposes that $k_+ = k_- = k_\pm$, where $k_\pm$ is the number of vortex pairs.

In Fig.~\ref{fig:vortex}, we show examples of vortices/anti-vortices in RBM-CosSin generated XY configurations at different temperatures. The existence of vortex pairs and their binding-to-unbinding change across the KT transition are well depicted in these machine generated configurations. 

\begin{figure}
\includegraphics[width=0.85\linewidth]{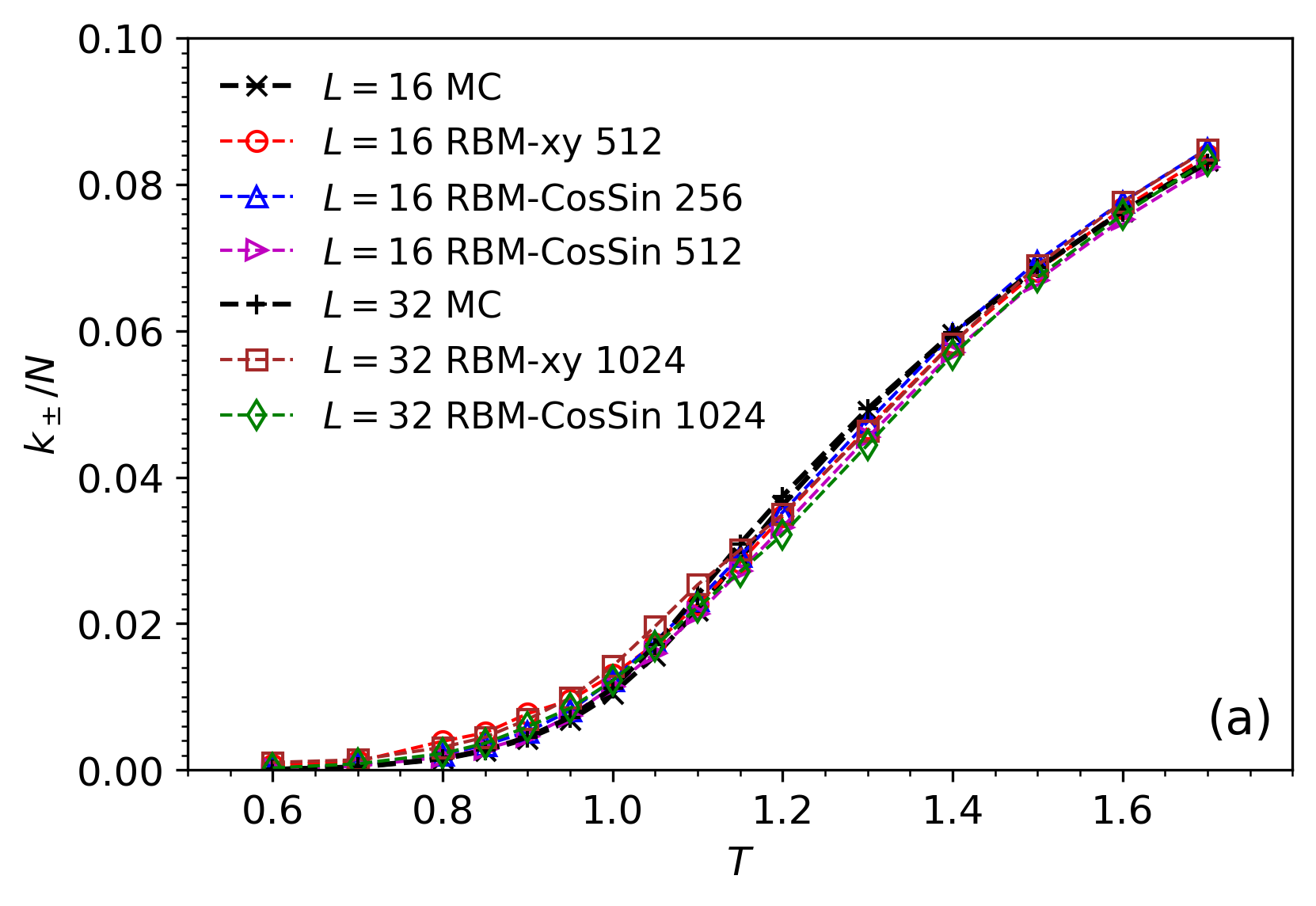}
\includegraphics[width=0.9\linewidth]{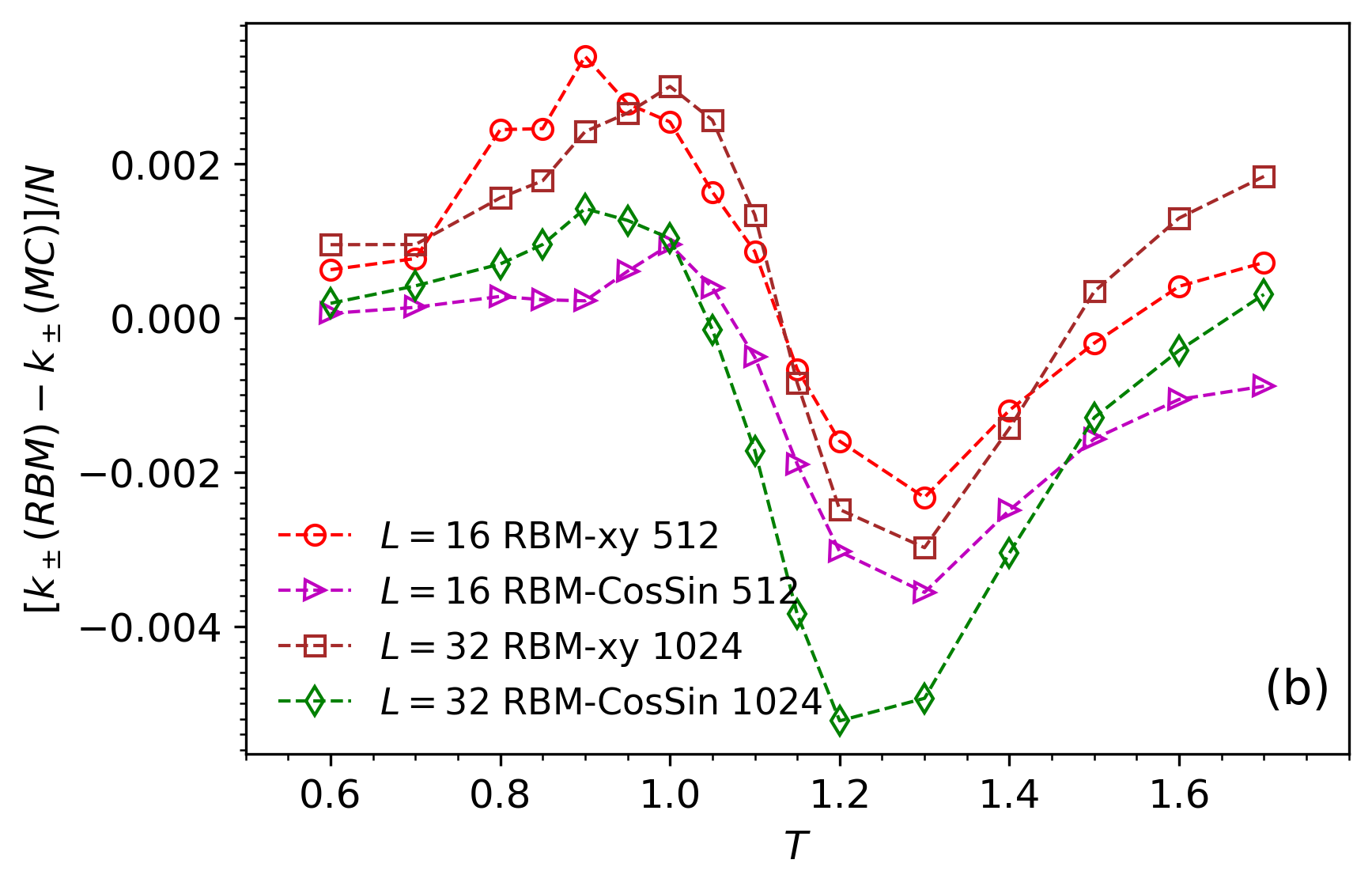}
\caption{\label{fig:wind_number} (a) Vortex pairs density ($k_\pm/N$) for RBM generated XY configurations at various temperatures as compared with Monte Carlo simulation results (MC) for systems of $L=16,32$.  (b) The difference between RBM generated and MC simulated $k_\pm$'s.  }
\end{figure}

To quantitatively evaluate the performance of RBM generated topological defects, we calculate the vortex pair density $k_\pm/N$ as a function of temperature, which is in remarkable agreement with Monte Carlo results (Fig.~\ref{fig:wind_number}a). Focusing on their difference,  RBMs tend to overestimate (underestimate) $k_\pm$ below (above) $T\approx 1.1$ (Fig.~\ref{fig:wind_number}b). It is well known that tracking the vortex number alone cannot detect the KT transition because the change of $k_\pm$ with temperature is gradual~\cite{ota1992} (although its changing rate does point to $T_p$~\cite{tobochnik1979}). Empirically, we observe a peak in the difference between RBM $k_\pm$ and Monte Carlo $k_\pm$ that aligns with $T_c$. This peak might be caused by the frustration of RBMs to capture topological defects during the KT transition, where unbinding of vortex pair occurs. However, even if this detection method is reliable, it still relies on the input of prior physics knowledge of what constitutes a vortex.

\subsection{Transition temperatures captured by RBM weight matrices}
\begin{figure}
\includegraphics[width=0.55\linewidth]{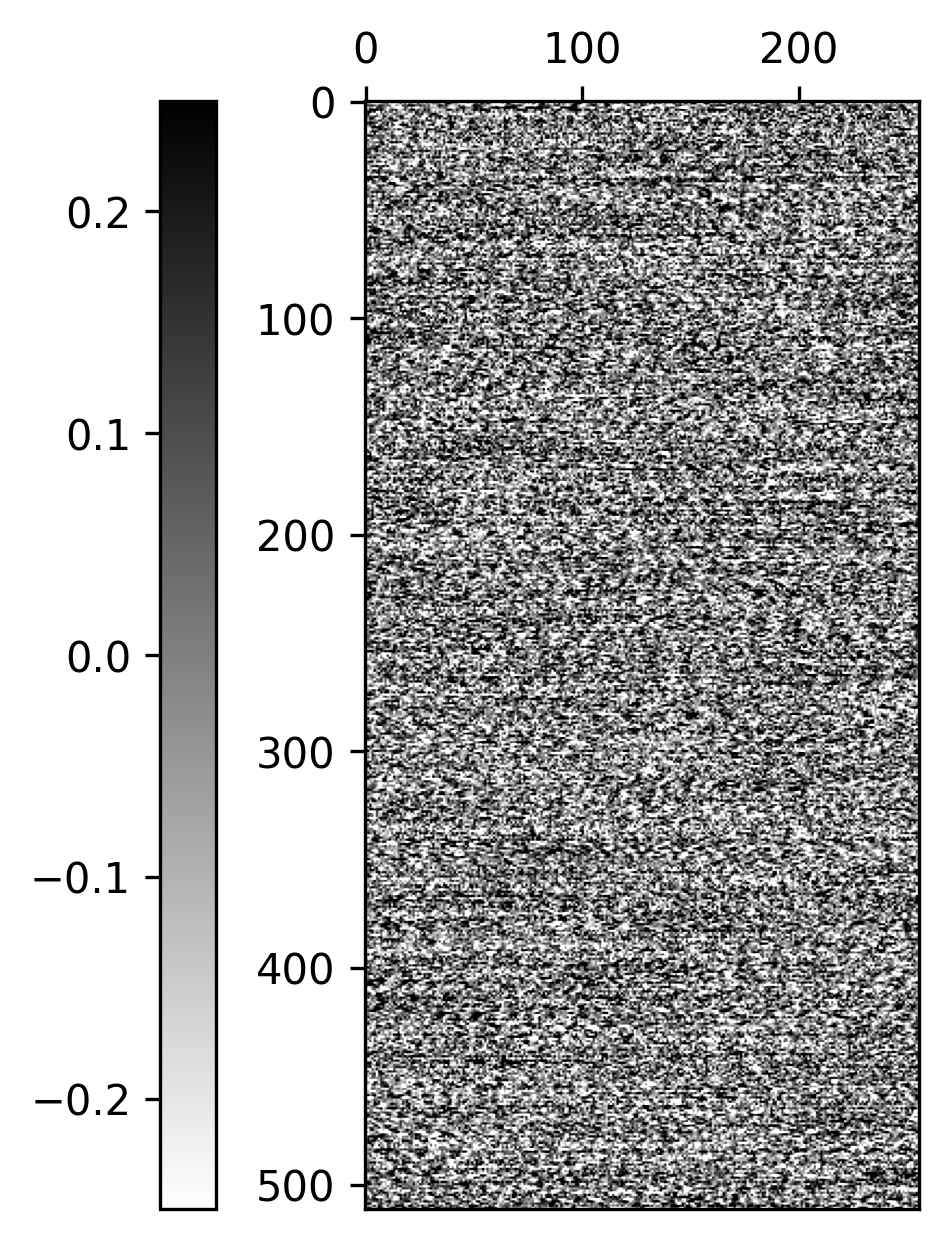}
\includegraphics[width=0.3\linewidth]{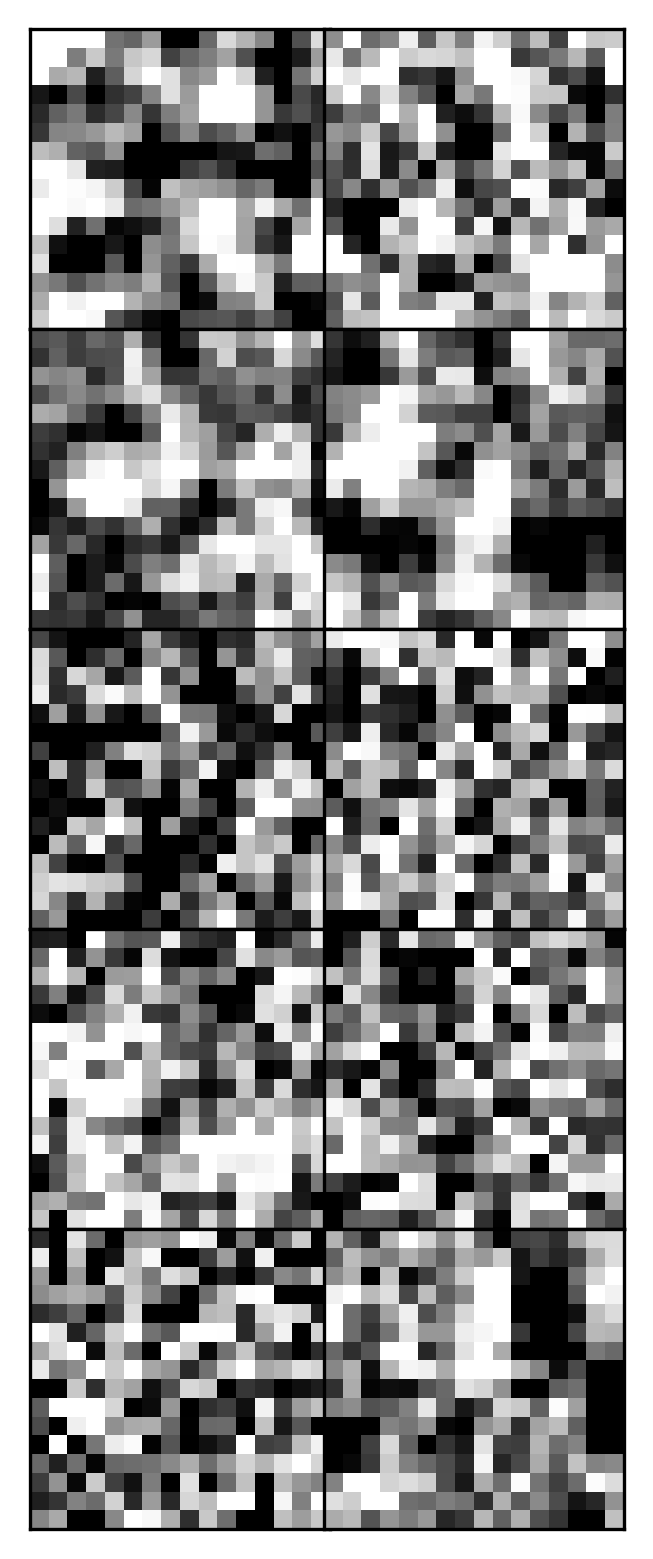}
\caption{\label{fig:Afilter}    $n_h\times n_v = 512 \times 216$ weight matrix ${\bf A}$ of RBM-CosSin at $T=0.6$.  Ten example filters ${\bf a}_i^T$ (rows) of the ${\bf A}$ matrix folded into $16\times 16$ matrices are shown on the right. Color bar is truncated between $[-0.25,0.25]$.  }
\end{figure}

It has been recognized that the weight matrices of RBMs contain rich information about the Ising phase transition~\cite{iso2018,cossu2019,gu2022}. Following the convention of this article, a row vector (e.g. ${\bf a}_i^T$) maps XY configurations onto a hidden unit $i$ (Fig.~\ref{fig:Afilter}). When folded into 2D $L \times L $ matrices, these filters exhibit patterns that match XY configurations. At a fixed temperature,  the collection of all $n_h$ filters reflects the variation of spin states of the corresponding Boltzmann distribution, which should capture thermodynamic fluctuations given by heat capacity or susceptibility. 

\begin{figure}
\includegraphics[width=0.9\linewidth]{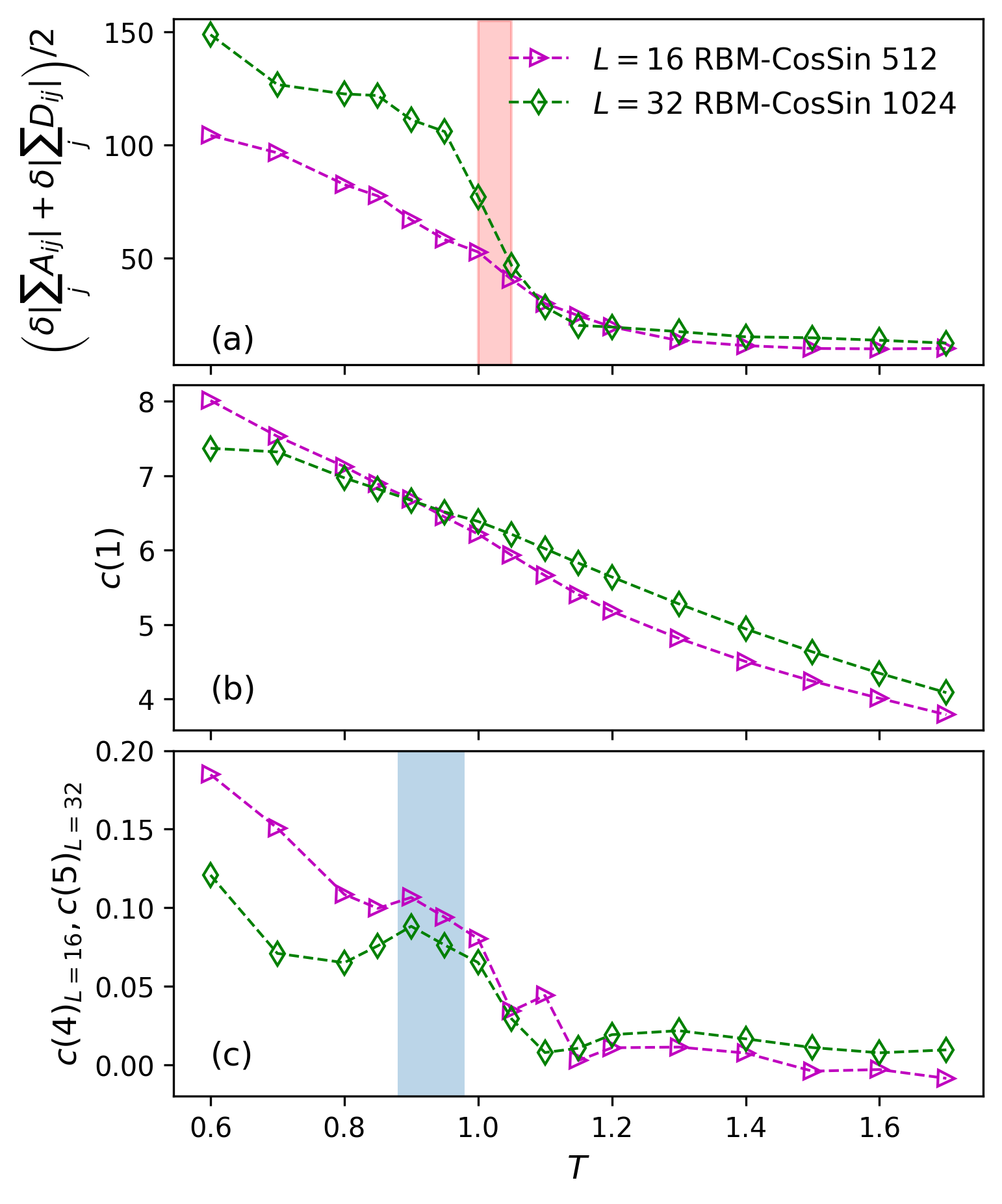}
\caption{\label{fig:AD_T}  (a) Filter sum fluctuation,  (b) nearest-neighbor spin-spin correlation, (c) fourth-nearest-neighbor ($L=16$) / fifth-nearest-neighbor ($L=32$) correlation  as a function of $T$ for two RBM-CosSin. $T_p$ and $T_c$ predicted by these metrics derived from weight matrices are highlighted by transparent red and blue rectangles, respectively.}
\end{figure}

We define the filter sum fluctuation as the variance of the absolute value of the column sum of weight matrix elements, i.e.
\begin{align}
\begin{aligned}
    \delta \left| \sum_j A_{ij} \right| & = \left\langle \left| \sum_j A_{ij} \right|^2 \right\rangle - \left\langle \left| \sum_j A_{ij} \right| \right\rangle^2 \\
        \delta \left| \sum_j D_{ij} \right| &= \left\langle \left| \sum_j D_{ij} \right|^2 \right\rangle - \left\langle \left| \sum_j D_{ij} \right| \right\rangle^2,
\end{aligned}
\end{align}
where $\langle \cdot \rangle$  is the average over all $n_h$ filters (rows).
We measure $\left(\delta \left| \sum_j A_{ij} \right| + \delta \left| \sum_j D_{ij} \right|\right)/2$ as a function of temperature (Fig.~\ref{fig:AD_T}a). Unlike the Ising model, for which filter sum fluctuation is low at both high and low temperatures resulting in a peak around its critical phase transition, the XY filter sum fluctuation decreases monotonically with temperature. This is because Ising ferromagnetic phases primarily oscillate between two states: spin up and spin down. The continuous symmetry and rotationally invariant nature of the XY model allows for a significantly larger number of possible states at low temperatures. Nevertheless, we observe a rapid drop of filter sum fluctuation around the heat capacity peak temperature $T_p$, especially in larger systems, which demonstrates RBMs' ability to capture this transition.

\begin{figure}
\includegraphics[width=0.85\linewidth]{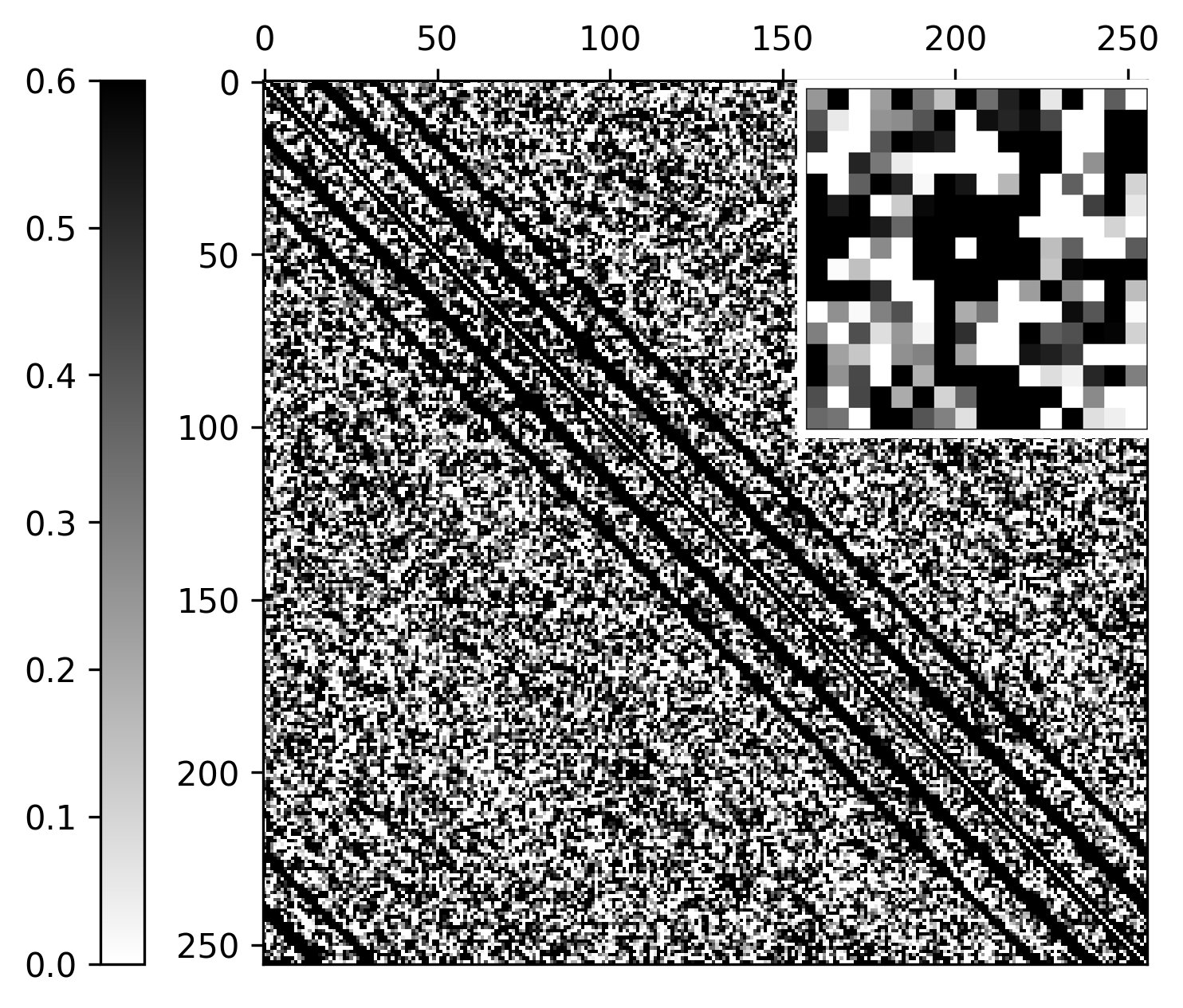}
\caption{\label{fig:ATA}  $256\times256$ ($L=16$) spin-spin correlation matrix ${\bf A}^T {\bf A}$ of $n_h=512$ RBM-CosSin at $T=0.6$. To enhance the visualisation of numerical values, the trivial self-correlation along the diagonal is set to zero and colorbar scale is set within $[0,0.5]$. Inset: the correlation ${\bf a}_{:, j}^T{\bf a}_{:,j'}$ between one spin $j=120$ placed at the center and all other spins $j'$ wrapped into a $16\times 16$ matrix. }
\end{figure}

Looking vertically, each column vector (e.g. ${\bf a}_{:,j}$) of the weight matrix is a mapping from hidden states to a spin $j$.  It is therefore expected that the inner product between two of these inverse filters, ${\bf a}_{:,j'}^T {\bf a}_{:,j}$, reveals the spin-spin correlation between the $(j, j')$ pair~\cite{iso2018}. We define the correlation between a spin $j$ and its $l$th-nearest-neighbors as
\begin{align}
c(l) = \frac{ \langle {\bf a}^T_{:,j} {\bf a}_{:, j+l} \rangle + \langle {\bf d}^T_{:,j} {\bf d}_{:, j+l} \rangle}{2}
\end{align}
where $\langle \cdot \rangle $ is the average over four neighboring sites at a distance $l<L/2$ and over all sites. It is more convenient to calculate the $n_v \times n_v$ correlation matrix, e.g. ${\bf A}^T {\bf A}$, whose $(i,j)$ entry is the correlation between the $(i,j)$ spin pair~\cite{cossu2019}. The correlation matrix of the XY model exhibits characteristic high value stripes parallel to the diagonal direction (Fig.~\ref{fig:ATA})~\cite{iso2018,cossu2019}. These dark strips correspond to strong correlations of a spin with its first, second, and third nearest neighbors, after properly folded into $L \times L$ lattices. When examined as a function of temperature, the first nearest neighbor correlation $c(1)$ decays monotonically and gradually,  lacking any distinct features to identify transitions(Fig.~\ref{fig:AD_T}b). 

In contrast to the physical picture of the KT theory that the (physics) correlation decays (as a function of distance) algebraically (exponentially) at fixed temperature below (above) $T_c$~\cite{kardar2007}, the RBM encoded correlation quickly drops to small values beyond  the 4-5th nearest neighbors, forbidding the extraction of a physical correlation length. In fact, the correlation $c(l)$ can be oscillatory (as a function of $l$) and experiences a local minimum around $l=4-5$ (Fig.~\ref{fig:ATA} inset). We expect this characteristic medium-range order around $l=4-5$ to be related to the length scale of vortex pair unbinding occurrence and the unbinding of vortex pair at $T_c$ could give rise to excess correlation over that medium distance. After plotting $c(4)$ or $c(5)$ as a function of temperature, we observe a rounded peak indicating the KT transition $T_c$ (Fig.~\ref{fig:AD_T}c).  This shows that we can extract useful metrics from weight matrix parameters of unsupervised RBMs for the detection of the topological phase transition without introducing physics knowledge into the method.   
 
\section{Conclusion}
\label{sec:conclusion}
In this work, we propose two types of real-valued RBMs that can effectively learn the underlying Boltzmann distribution of the 2D XY model. In particular, the second type (RBM-CosSin) can generate new configurations that genuinely resemble true XY configurations by accurately capturing the thermodynamic properties and topological order of the physical system. Besides measurements based on physical concepts, such as the divergence of susceptibility and vortex pair density, we find that the weight matrix parameters of our RBMs encode important information that can be harnessed to predict the topological KT transition $T_c$ and the higher crossover temperature $T_p$. The success of RBMs with such simple architectures in tackling this long-standing task might be attributed to the similarity between energy-based models and statistical physics systems, as well as the suitability of the unimodal von Mises distribution for angle sampling. It would be interesting to generalize the current model and method to study other vector field systems, such as liquid crystals~\cite{lebwohl1972}, active matters~\cite{vliegenthart2020}, and 
biological tissue defects~\cite{killeen2024}.


\begin{acknowledgments}
We acknowledge the UT Tyler startup fund, the Welch Foundation and the Texas Advanced Computing Center (TACC) at the University of Texas at Austin for supporting this work. We also thank Jing Gu for correcting a bug in the earlier version of the code.
\end{acknowledgments}



%

\end{document}